\documentclass[12pt]{JHEP3}
\usepackage{amsfonts,amssymb}
\usepackage{amsmath}
\usepackage{epsf}

\def\T11{{T}^{1,1}}
\def\bear{\begin{eqnarray}}
\def\eear{\end{eqnarray}}

\newcommand{\vac}{{|0\rangle}}

\newcommand{\pa}{\partial}
\newcommand{\tr}{{\rm tr}}
\newcommand{\comment}[1]{}
\newcommand{\CO}{{\cal O}}
\newcommand{\pasl}{\pa\kern-.55em /}

\newcommand{\ksl}{k\kern-.55em /}

\newcommand{\ket}[1]{|#1\rangle}

\newcommand{\braket}[2]{\langle #1|#2\rangle}

\DeclareFixedFont{\xiiss}{OT1}{cmss}{m}{n}{12}
\DeclareFixedFont{\ixss}{OT1}{cmss}{m}{n}{9}
\DeclareFixedFont{\cmrnine}{OT1}{cmr}{m}{n}{9}
\newcommand{\field}[1]{\mathbb{#1}}

\newcommand{\BC}{{\field C}}
\newcommand{\BR}{{\field R}}

\newcommand{\CCs}{\hbox{\ixss C\kern-.4emI}}
\newcommand{\ZZs}{\hbox{\ixss Z\kern-.4emZ}}

\newcommand{\CP}{{\BC\field P}}

\newcommand{\expec}[1]{\langle #1\rangle}

\newcommand{\myfig}[3]{\begin{figure}[ht]
\begin{center}
\leavevmode
\epsfxsize=#2cm
\epsfbox{#1}
\end{center}
\caption{#3}
\label{fig:#1}
\end{figure}}

\title{Large $N$ BPS states and emergent quantum gravity}
\author{David Berenstein\\ 
Department of Physics, UCSB, Santa Barbara, CA 93106\\
Email: \email{dberens@physics.ucsb.edu}}
\abstract{ This paper provides a heuristic derivation of how classical gravitational physics  in the AdS/CFT correspondence appears from the strong dynamics of the ${\cal N}=4$ SYM theory in a systematic way.
We do this in a minisuperspace approximation by studying $1/8$ BPS configurations.  We can show that our description matches the semiclassical physics of $1/8$ BPS states in supergravity. We also provide a heuristic description of how massive strings appear in the geometry, and how at strong 't Hooft coupling they become local on the $S^5$ suggesting that they can be realized as a sigma model on a weakly curved background.
We show that the dynamics of $1/8$ BPS dynamics of ${\cal N}=4 $ SYM on a round $S^3$ can be reduced to that of a matrix model for commuting 
matrices. Including measure factors, we show that this effective dynamics is related to bosons
living on a six dimensional phase space with repulsive interactions. Because of these interactions, we
can argue that on the ground state the bosons assemble themselves on a spherical shell in the shape of a round five sphere. This sphere will be identified with the $S^5$ in the AdS dual geometry. To do this, we first define a precise way to coarse grain the dynamics. We use
half BPS configurations as a toy model for this coarse graining, and we can reproduce the
droplet picture of these half BPS states systematically. The droplet appears as the saddle point approximation of a statistical ensemble related to the square of the wave function of the eigenvalues of a complex matrix. This procedure is also applied to the set of $1/8$ BPS configurations to extract the geometry, giving an analog of the droplet picture of half BPS states for the case of  $1/8$ BPS configurations. We also have a conjectured realization of some $1/8$ BPS giant
graviton wave functions in the dynamics, which  captures all $1/8$ BPS giant gravitons constructed by Mikhailov.  This leads to a lot of different topology changes which can be treated heuristically.}

\preprint{hep-th/0507203}

\begin{document}

\section{Introduction}

One of the most difficult problems in theoretical physics is to solve strongly coupled 
field theories and understand their effective dynamics. For example, take the theory of the 
strong interactions, QCD \cite{QCD}. It is expected that the theory confines, and that at low energies the effective dynamics of the theory reduces to a collection of mesons and perhaps glueballs. Taking only the least massive fields (the pions), we get what is called a chiral lagrangian.
 These degrees of freedom are not apparent in the UV description of QCD, in terms of a non abelian gauge dynamics for the gauge group $SU(3)$ and the massive quarks. 
This low energy dynamics is usually claimed to be an {\em emergent phenomenon} of the theory of strong interactions. 

The analysis of the field theory is improved with the idea of 't Hooft of the large $N$ expansion\cite{largeN}. The claim of the large $N$ expansion is that in the large $N$ limit, QCD is well described by some type of weakly coupled string theory. The mesons become 
open strings, and the glueballs are closed strings. There is some scale associated to the string dynamics (the effective string tension), and the string coupling constant is of order $1/N$
at that scale. This argument is done at the level of perturbation theory. The claim is that
this analysis can be extrapolated to the strong coupling regime. In this sense, the theory is said to have a large $N$ limit if all of the statements above are true as an asymptotic expansion
around $N=\infty$. In particular, at $N=\infty$ we seem to get a free theory of mesons and glueballs
(a free string theory), a property which should make the theory simple to understand (a very lucid discussion can be found in \cite{Witbaryons}).

In this paper, we want to understand what does it mean when we speak of gravity as an emergent phenomenon in a quantum system, in the same sense as the chiral lagrangian is an emergent phenomenon in QCD. Indeed, this is some of the content of what is meant by claiming that in the AdS/CFT correspondence \cite{M},  the CFT is a definition of quantum gravity (as defined by type II strings) on an $AdS$ geometry. In this paper we will be concerned only with type $IIB$ string theory on $AdS_5\times S^5$ and it's dual ${\cal N}=4 $ SYM theory in four dimensions, which is the most well studied example of the correspondence. The SYM theory is also a large $N$ theory in four dimensions and fits into the scheme of 't Hooft for the large $N$ expansion of the field theory being described by a string theory. The fact that these two seemingly different theories can be related is due to the holographic nature of gravity \cite{Holography}, whereby the total number of degrees of freedom in the gravity side is not an extensive quantity. The notion of holography is in general imprecise. We can  have bounds on the numbers  of degrees of freedom associated to a region of spacetime, but we do not seem to have a microscopic understanding of these bounds. The most covariant bounds one can write are given by the study of light sheets in the geometry by Bousso \cite{Bousso}. 

The analogy to QCD should not be lost here. The difference between the AdS/CFT correspondence and QCD, is that only in the first case do we have a precise identification of a semiclassical (large volume) ten dimensional geometry where the strings propagate when we let $N$ become very large. Moreover, the strings that appear are critical strings which can be quantized on flat space. Given this fact, we can try to do a systematic expansion around flat space to extract the dynamical content of the theory. This is what lets us do order of magnitude estimates for string energies, and semiclassical gravity calculations.
A lot of work has ben done in trying to understand the dual string theory to QCD, but success so far has been limited, and the only thing we seem to know for certain is that the QCD string lives in at least five dimensions \cite{Polyakov}. 

This paper is an attempt to derive the strong coupling dynamics of the ${\cal N}=4 $ SYM 
CFT from first principles. Because we can not do this in full generality, we will restrict our attention to supersymmetric states.  We will try to show that close to supersymmetric states, this dynamics is given by studying local objects on an auxiliary geometry, which we will relate to a piece of the $AdS_5\times S^5$ geometry, or some other geometry with asymptotic $AdS_5\times S^5$
boundary conditions.

We will attempt to prove that locality in the $S^5$ follows from doing both a coarse graining approximation to the dynamics: by looking at supersymmetric states and describing the states in terms of collective coordinates, and from taking the SYM 
theory to strong coupling. Non-BPS excitations will then be added as  a perturbation of the BPS
configurations. 
Our approach will be based on a self-consistent approximation to the dynamics of BPS states in the large 't Hooft limit. The results of this paper are not claimed to be exact. They are of a qualitative nature. However, the picture presented by these arguments is very compelling. In the end, we can only claim that we have derived a qualitative picture of locality on the $S^5$. Together with the time and the $S^3$ on the boundary, one can reconstruct locality in nine out of ten directions of the $AdS$ geometry.
This is very suggestive that eventually the full geometry of $AdS_5\times S^5$ can be derived from first principles.

The paper is organized as follows:

%To begin, we write a preliminary section \ref{sec:prelim} which can be construed
%as a philosophical discussion of the ideas that we are using in this paper. This section can be skipped on a first reading of the paper. It serves to establish the meaning of some of  the language that is used, and 
%to justify some of the choices that are done in the main body of the paper. It also discusses
%many issues necessary for the interpretation of some of the results in this paper..

 In section \ref{sec:halfbps} we study the dynamics of half BPS states from a new point of view. We show that in order to study the system of half BPS states it is necessary to go beyond counting the degeneracy of states, and one needs to study rather an operator algebra on the Hilbert space of states.
We also show by using well known arguments that have been used in the study of the quantum hall effect that the notion of the shape of the droplet can have a very precise meaning. This proceeds by first coarse graining the system, and describing the dynamics in terms of a density of eigenvalues in the quantum plane. One can relate this picture to a non-relativistic Coulomb gas in two dimensions by looking at the square of the wave function of the system as a Boltzman partition function. The droplet appears from the saddle point approximation of this partition function in terms of coarse grained variables, given by the density of eigenvalues on the quantum plane.
We show this way that the shape deformations of the droplet 
are obtained directly from coherent states of the duals of gravitons in the CFT description in the same way. This approach clarifies many statements that have been made in the literature before.

In section \ref{sec:chring} we study the set of $1/4$ and $1/8$ BPS operators/states  in the SYM theory and we show that using the operator state correspondence they should be associated to a matrix model with finitely many matrices which commute. In the case of $1/8$ BPS operators we show that they are related to a first order dynamics of $3$ complex matrices and two fermionic matrices. We also show by counting states that there are no $1/8$ BPS black holes in $AdS_5\times S^5$.

Next, in section \ref{sec:matrix} we study in detail how to extract the dynamics of the matrix model for commuting matrices by doing a careful semiclassical calculation. This section might be skipped on a first reading.
We show that for bosonic matrices, the system becomes a set of bosons in six dimensions in the presence of a strong magnetic field, with repulsive two body interactions, and also with three body interactions.
We present strong, but not definitive arguments that let us write the general wave functions of the system in terms of a holomorphic quantization of symmetric polynomials, times some measure factor which we calculate using the leading semiclassical approximation. This is akin to describing the half BPS states in terms of all free fermion wave functions for $N$ fermions in the harmonic oscillator. 

In section \ref{sec:saddle} we use the same techniques used in coarse graining the dynamics of the half BPS state for the case of $1/8$ BPS states. We show that the ground state can be described as some statistical Boltzman gas of particles in a harmonic oscillator with repulsive logarithmic interactions in six dimensions. This is not a Coulomb gas problem, and  needs to be solved for. This in general leads to a set of integral equations for the density of particles that need to be solved. Using some functional properties of the two body potential we show that
the distributions of particles is singular and all the particles are uniformly distributed on a round five sphere. This is like a five dimensional membrane, an eigenvalue 5-brane. Because we will be dealing with D-brane states as well, we will call this membrane an $E$-brane. The size of the E-brane is of order $\sqrt N$, which is large compared to $\hbar$ and predicts that the system can be understood classically.
 We can also use coherent states to find that the graviton coherent state wave functions deform the shape of this E-brane exactly as one would expect from identifying the $S^5$
 shaped E-brane  with the $S^5$ in the supergravity description, and we can identify the speed of sound on the E-brane with the speed of light on the $AdS$ geometry.

In section \ref{sec:top} we discuss topology changes of the E-brane by nucleating giant gravitons. We identify individual particles away from the E-brane with giant graviton 
D-branes expanding into the $AdS_5$. We also conjecture the form of the giant graviton wave function growing into the $S^5$ by using holormorphy and the half BPS giant graviton wave functions as an example in terms of determinants of holomorphic functions. 
We show that the main effect of these wave functions is to repel eigenvalues from the intersection locus of the zero set of the holomorphic functions and the $S^5$. This suggests that their shape in $S^5$ is given exactly by such an intersection, matching the characterization of giant gravitons by Mikhailov \cite{Mikh}.

In section \ref{sec:string} we describe heuristically the set of states that correspond to strings propagating on the E-brane. We show that as the 't Hooft coupling constant becomes strong, these are better described by a curve which is tangent to the E-brane, suggesting that the
strings probe the E-brane geometry locally and via a sigma model type action. 

We close the paper with a discussion of the results found in this paper and an outlook of future directions that should be explored.

\section{The dynamics  of 1/2 BPS states }\label{sec:halfbps}

In this section we consider the problem of understanding the dynamics of the $1/2$ BPS operators/states in the ${\cal N}=4 $ SYM theory. This should be considered as a 
first step to understand the main problem we are considering in this paper. This is also done to 
explain the criteria that are going to be used later on in the paper to determine how successful we have been in describing the more complicated problem of $1/4$ and $1/8$ BPS operators.

Let us begin with an analysis of the half BPS operators in
${\cal N}=4$ SYM.
It turns out all of these operators are very simple.
They are elements of a representation of the superconformal group. These are classified by a highest weight state in the $SU(2,2|4)$ representation (which labels the half BPS multiplet). This highest weight state is called a (super)-primary operator, and all other states are 
called descendants. The primary state is
built out  of a single complex scalar $Z$. The complex scalar field $Z$ is unique once we specify which half of the supersymmetries are to be preserved by the state.

These half BPS primaries are generically multi-trace operators of the form
\begin{equation}
\prod \tr(Z^{n_1}) \dots \tr (Z^{n_k})
\end{equation}
with $N\geq n_1\geq n_2 \dots \geq n_k$ \cite{CJR}. The ordering appears because different traces commute, so their order in the operator  does not matter. How many $Z$ are bundled in each  trace does matter.
The bound of $N$ for the $n_i$ appears  because $Z$ is an $N\times N$ matrix,  and there are identities (Cayley-Hamilton identities) 
that let us write $\tr(Z^{N+k})$ in terms of smaller numbers of traces.
Because of this inequality for the $n_i$, one can associate to this operator a Young 
tableaux (meaning an arrangement of boxes in a corner satisfying the constraints of Young tableaux) with columns of length $n_1, \dots , n_k$. The operator has a  conformal weight equal to 
$\sum_i n_i$ (the number of boxes). This counting captures 
all of the 1/2 BPS states of the system \cite{CJR}.  At this moment we could stop and say we have solved the $1/2$ BPS dynamics. But in view of the points made previously in the introduction, we have only solved the energy problem: how many states we have at each energy, without any additional structure telling us how they are actually related. Already this can be useful to count entropy or to calculate a specific heat, but it is not enough for what we want.

To solve the dynamics, we need to go further, and consider what are the natural operators
in our Hilbert space of states, and how to calculate their expectation values or matrix elements in this basis.
In a conformal field theory, it is natural to consider OPE coefficients in an orthogonal basis. Unfortunately, the basis of traces is not orthogonal, although for fixed energy of order one it is approximately so in the large $N$ limit, and the failure of orthogonality is of order $1/N^2$.
However, we want to consider situations where the total energy can be very large as well (let us say of order $N^2$) where the planar approximation usually breaks down. 
In fact, the failure of orthogonality of traces starts to become of order one when the energy $E$ is
of order $\sqrt N$ \cite{J2N,BN}, which is much smaller than the size of the matrices. This was discovered by carrying out the planar expansion of free diagrams in the BMN limit \cite{BMN}, where large values of $E$
became important to describe rapidly moving strings in $AdS_5\times S^5$. 

Because ${\cal N}=4$ SYM theory is a conformal theory, to every such operator there is an associated state in the theory, when the theory is  compactified on a round $S^3$.
Being careful about the operator state correspondence, one finds that the states associated to these operators are built out of traces of the s-wave scalar mode of the field $Z$ on a round $S^3$ (but not it's complex conjugate). Because of this fact, the effective dynamics describing all of these states 
reduces to the problem of studying a matrix model quantum mechanics for a single Hermitian matrix, which was identified with the large $N$ harmonic oscillator gauged quantum mechanics \cite{toyads}. 
This model is solvable, and leads to a theory of free fermions in the one-dimensional harmonic oscillator potential\cite{BIPZ}. It can also be interpreted in phase space in terms of a quantum hall droplet for free fermions in the lowest Landau level.  {
To do this, it is necessary to twist the Hamiltonian $\Delta$ for the field theory on $S^3$ (the Hamiltonian is related to the generator of dilatations in the Euclidean field theory ) to 
\begin{equation}
H = \Delta -J
\end{equation}
where $J$ is one of the generators of the $SO(6)$ R-symmetry charge. For the Hamiltonian $H$, all half BPS states satisfy $H=0$, and all other states in the theory have $H\geq 1$ perturbatively. Because of this property, it is in principle possible to take a decoupling limit where all
states with energy $H>\epsilon$ are forbidden. We can then do a reduction of the dynamics to only those states which satisfy $H=0$ and  write an effective dynamics which captures all of these states. This is a reduction to the degeneracies of the ground state, like the Lowest Landau Level problem for a particle in a magnetic field. However, here the dynamics for the lowest Landau level is for $N$ identical particles.}

What we want to describe now is how do we determine the precise details of the dynamics for these states, when the naive prescription for finding states by traces does not work well for us, as it does not produce orthogonal states. The idea is to step back for a second and build by hand a conjectured dynamical system that reproduces the correct counting of states. It turns out that the final answer is not obvious: there are many systems whose energies are given exactly by that same counting of states, so that finding the spectrum of the Hamiltonian alone is not a sufficient criterion for understanding the microscopic details of the system. We want to elaborate further on this point by setting an example of two systems that produce the desired dynamics.

For example, we can consider a system with $N$ identical bosons on a harmonic oscillator.
This can result from making a diagonal ansatz for the field $Z$, where only the $s$-wave of the field 
$Z$ is turned on $S^3$. Classically, we get exactly this system of $N$ bosons in this 
subsector. Indeed, this is up to gauge invariance the most general classical configuration which has $H=0$ \footnote{The reader might be confused by the fact that we are only talking of turning on the field  $Z$ and not $\bar Z$. In the free field theory, $Z$ has two pieces with different time dependence, one that creates quanta for $Z$ and one that destroys quanta for $\bar Z$. The proper statement is that we are setting the system to the vacuum state for the
$\bar Z$ quanta, and keeping the state for the $Z$ quanta to be general. This reduces the degrees of freedom to one quantum per eigenvalue of $Z$. The apparent doubling of degrees of freedom by using a complex variable,  is that the real and imaginary piece of $Z$ encode the information about a canonically conjugate pair of variables. Also once this reduction is done, $Z$ commutes with it's complex conjugate because of the Gauss' constraint. So the model 
corresponds to first order dynamics for a normal matrix.} 
These are bosons because once we choose a diagonal form for $Z$, we can still permute the eigenvalues by elements of the $SU(N)$ gauge group,  which gives us a discrete group of permutations 
$S_N$. Gauge invariance requires these states related by permuting the eigenvalues to be equivalent, and hence the wave functions should be totally symmetric in the eigenvalues. We will take this dynamics as our first candidate dynamics.

This system of $N$ bosons will give the same counting of states as above. 
The bosons can be organized by energies according to  Young Tableaux as well, so that the
boson with the highest energy has as much energy as the first row of the Tableaux, the one with the second highest energy has as much energy as the second row of the tableaux, etc. The boson statistics tells us that this is all we need to do to describe this system.

The ground state wave function for these bosons can be given as follows (in a complex basis for the phase space $x, p$, $z= x+ip$)
\begin{equation}
\vac=  c \exp( -\sum \frac 12 z_i \bar z_i)
\end{equation}
where $c$ is a normalization factor so that $\braket{0}{0}=1$. This puts all of the bosons in the ground state of the harmonic oscillator. All of them are sitting classically at the origin, and we have a delta function
distribution of the boson density in the semi-classical limit.
To excite the fist eigenvalue  to the energy $r_1$ over the ground state we multiply the above by $z_1^{r_1}$, and to excite multiple bosons, we multiply by 
\begin{equation}
\ket{r_1, r_2, \dots}= z_1^{r_1} z_2^{r_2} \dots \vac
\end{equation}
However, here we are being careless. We need to make sure that the wave function is symmetric under the exchange of eigenvalues. 
To do this, we symmetrize the wave functions by summing over the different choices for the eigenvalues. 
For example
\begin{equation}
\ket{r_1}= \sum z_i^{r_1}\vac
\end{equation}
and
\begin{equation}
\ket{r_1, r_2} = \sum_{i\neq j} z_i^{r_1} z_j^{r_2}\vac
\end{equation}
We see this way that we can describe a system which has the same energies (above the ground state) as
the system we are trying to describe.
Also, one can consider the trace basis as consisting of multiplying any state by the symmetric function $\tr(Z^j) = \sum_j z^j$.

This seems to be a perfectly reasonable realization of the system,  as long as we ignore
$1/N$ effects and work at finite energy. Under these conditions it seems as  if we have the  same dynamics as before,
with multiple traces creating approximately orthogonal states \footnote{One can check that  overlaps are of order $1/N$ in this case, and not $1/N^2$, but if we are ignoring subleading corrections, the systems agree}. In this case, each trace
acts like a particular raising operator for an oscillator with frequency $n$, and the traces get truncated at order $N$.

It has also been shown that one can describe the quantum system also in terms of $N$ fermions in the harmonic oscillator, as opposed to $N$ bosons (see for example \cite{toyads}). 
The energy levels above the ground state of the system of the fermions are identical to those of the system of bosons. The picture in terms of fermions in phase space can be connected to 
a quantum hall droplet sample, and gives rise to an incompressible quantum liquid system.
The most naive way to  differentiate between the fermionic and the bosonic system is that
the wave functions differ by the Vandermonde determinant, namely
\begin{equation}
\ket{r_1, \dots, r_i}_F \sim \ket{r_1, \dots r_i}_B \prod_{i<j}(z_i-z_j)
\end{equation}
where the subscripts $F,B$ refer to the Fermion and Boson wave functions for the states. 
Because of Fermi statistics, the fermions can not all be placed at the origin. Instead, the ground state is determined by the Fermi sea level, which in this case is a circle centered around the origin of radius of order $\sqrt N$ in units of $\hbar$. The pictorial description of the ground states is encoded in
figure \ref{fig:BvsF.eps}. 

\myfig{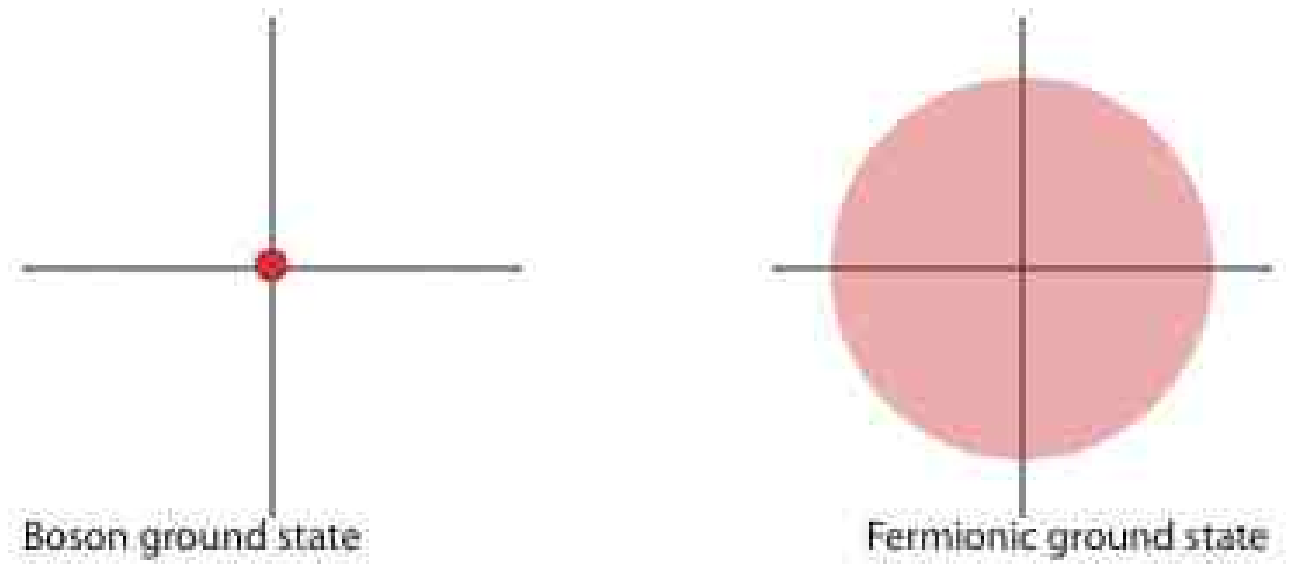}{6}{Pictorial comparison of the boson and fermion ground state wave functions}

Again, in the fermion wave function, multiplying by traces generates the set of all fermion wave functions, and we get approximately orthogonal states. In the fermionic case, the orthogonality becomes of order $1/N^2$, and in fact it coincides with the matrix model planar diagram expansion. However, we need to remember that the matrix model is derived in the weak coupling approximation of SYM. At strong coupling we need to worry that 
the analysis done at weak coupling is not spoiled.

The question is now to decide which is the correct dynamics of the system, both of 
which seem relatively simple and both of which count the states correctly. 
 We can decide this question  by looking at the 
supergravity duals of $1/2$ BPS states.

The results of Lin, Lunin and Maldacena from the supergravity approach \cite{LLM},  show that at the  supergravity level, the dual objects to these states with smooth supergravity solutions correspond to geometries determined by black and white paintings of the plane, with the total area painted in black being equal to $N$ in appropriate units. This is interpreted as flux quantization in supergravity. This picture is compatible with
having the black patches filled with an incompressible (quantum) liquid. This picture selects the quantum hall droplet description over the bosonic counterpart.
Still, it is not obvious that one should select free fermions in the lowest Landau level, and not a fractional quantum hall liquid instead, or some other dynamical system with similar properties. After all, in the limit where supergravity is valid,  one  is effectively at strong coupling in the SYM dual. \footnote{One can evade this type of consideration  by using strongly the fact that the full dynamics of these special types of objects are protected by supersymmetry and therefore should coincide with the perturbative description of the system. This requires a non-renormalization theorem for the effective action. Although the author believes this is true, it is not clear in this context
how one would apply the non-renormalization theorems which were proved for flat space supersymmetry, which has a different group structure than the superconformal groups.}

Given this fact, one has to wonder why does the supergravity picture select the fermion
droplet as the correct description of the system, while the bosonic picture seems just as 
good at this level.
What we will see now is that although it is true that in some sense the quantum systems are equivalent (in that the spectrum of the Hamiltonian is the same in the  two systems), it is not true that they coincide as quantum systems. They really represent different   dynamical systems, even though they become identical in the strict classical level $\hbar=0$.

To test the difference in dynamics, we can try to understand the difference in normalization between the states created by traces, for example
$\ket{m}_B$ and $\ket{m}_F$, which are given by $\tr(Z^m)\vac_{B,F}$, which seem to be naively related to each other.
It is easy to show that 
\begin{equation}
\braket{m}{m}_B = N \frac{\int d^2 z \exp(-z\bar z) (z\bar z)^m}{\int d^2 z \exp(-z\bar z)}
\end{equation}
so that it scales uniformly with $m$ like $N$.

Now, let us consider the same problem in the fermion basis.
The Van-Der-Monde determinant is an alternating sum over permutations of the form
\begin{equation}
\sum_{\sigma} (-1)^{|\sigma|}\prod_{k=1}^N  z_{\sigma(k)}^{N-k}  
\end{equation}
If we multiply by $\sum z_i^m$ many terms can cancel: those for which we get repeated exponents between two of the $z_i$.
 
 One can then show that 
\begin{equation}
 \braket{m}{m}_F = \sum_{k=1}^m 
 \frac{\int d^2 z \exp(-z\bar z) (z\bar z)^{m+N-k}}{\int d^2 z\exp(-z\bar z)(z\bar z)^{N-k}}
\end{equation}
This later  sum looks  like
\begin{equation}
 \braket{m}{m}_B \sim \sum_{k=1}^m 
 \frac{(N+m-k)!}{(N-k)!} = \frac 1{m+1}\left(\frac {\Gamma(N+m+1)}{\Gamma(N)} -\frac{\Gamma(N+1)}{\Gamma(N-m)}\right) 
\end{equation}
which for small $m$ scales like $m N^m$. This is just right to coincide with the planar diagram expansion of the double trace correlator in the matrix model \cite{J2N}.

It follows that although the systems look naively equivalent (in that their energy spectrum coincides), their natural operator algebras are quite different, since they produce different normalizations for the states after using operators that look similar. One can generalize this further to states built by double traces acting on the ground state.
 Here, even if one changes the normalization of the operators $\tr(Z^M)$, these changes are 
 not enough to make them equivalent on the Hilbert space of states, and there is an honest distinction between the two systems. 
The second one,  given by the fermions, actually coincides with the description that arises from gauge field theory. This also coincides with the description obtained from supergravity correlators \cite{LMRS}. The key distinction between the naive model of bosons and the one with fermions comes from the fact that in the original derivation of the fermion  model, the volume of the gauge orbit was taken care of appropriately, while in the boson system this was ignored. This is, the gauge degrees of freedom are accounted for properly.
This volume is exactly the square of the Van-der-Monde determinant. One then absorbs a square root of this volume in the wave function of the system to obtain the fermion dynamics.
 This makes the measure on
each eigenvalue standard at the price of doing a similarity transformation in the quantum mechanical system.  Due to some miraculous cancellations, this produces free fermions in the end \cite{BIPZ}.
Classically, the fermion and boson system can not be distinguished, but they are different quantum mechanically. It is also possible to make a map between the states of both systems and impose by hand some identification of the operators between them. However, natural operators in one representation look very unnatural after we do this map. This is in the end the only sense in which the dynamics is truly different:
if we consider the full set of operators on the Hilbert space of states, which has a special basis adapted to the energy operator, the two systems are formally equivalent. Their distinction becomes apparent only when we require a finer structure to be available for the full description.
This is provided by a preferred coordinate system on phase space with respect to which we do our quantization and description.

One can make other tests that display their difference. For example, one can consider the ground state
expectation values  $\expec{\sum_i x_i^n}$ in phase space, in units of $\hbar$. In the boson case, this scales as $N$, while in the fermion case
it scales as $N^{n/2+1}$. This last dependence on the number of particles $N$ determines the natural 
description of the system from the gauge field theory point of view: this is the same one we obtain from doing planar diagrams in the ungauged theory, where we replace the sum above, by $\expec{\tr(X^n)}$, for $X$ a hermitian matrix. 

In the end, the fermion description 
provides for us a microscopic description of the supergravity droplets, although we have not yet explained why and how classical physics becomes part of the description. Part of the objective of this paper is to understand exactly what type of  finer structure is required to describe the quantum dynamics of the AdS/CFT correspondence so that classical physics  becomes possible. This classical behavior of the system is where spacetime geometry will ultimately come from.

For the half BPS states, the individual fermions are interpreted as D-branes (giant gravitons growing into AdS), and the droplets are formed by condensing various of these branes on top of each other (to the extent allowed by Fermi statistics). The rationale for studying these particular configurations is that large collections of stacked D-branes should have a weakly curved supergravity dual geometry. This reasoning also applies to coherent states of geometric fluctuations of these geometries. This last part will end up providing the shape moduli of the droplets. This type of description in terms of droplets and their shape is determined by collective effects of the dynamics, as most of the fermions play no significant role other than as filler of the droplets. The dynamics of shape is on the edges between the full and empty regions of the plane.
 Because of this behavior, the description in terms of droplet shapes should be considered as both a thermodynamic description (requiring large numbers of fermions to make the shapes sharp), and as emergent phenomena (the collective dynamics of the fermions can be matched to the geometry of spacetime in a controlled way as we take $N\to \infty$). We will explain this in what follows.
 
\subsection{Thermodynamic description of fermion droplets}

Let us consider the system of free fermions in a two dimensional 
phase space, in the presence of a quadratic potential given by $x^2/2+p^2/2$. The ground state of the system is going to be a circular droplet in phase space. If the number of fermions is $N$, the radius of the droplet is going to be of order $N^{1/2}$. The reader should notice that this description implicitly contains a lot of information. Because of our context, we know that the droplet is made of an {\em incompressible quantum fluid}, the quantum hall liquid. The density of the fluid is constant and determined exclusively by the quantum of area $\hbar$, as we are working in phase space. This description is coarse-grained: we are introducing the concepts of density on phase space, and order between the degrees of freedom, so that the wave function
 is entirely determined by a "classical" cartoon. Quantum mechanically, we know that in principle all possible configurations on phase space are valid with some probability density determined by the wave function. Why should we pick this one over all others?
We want to understand precisely how 
this description follows in detail from a microscopic description of the system via some mathematical
formulation. This is what we will do in this section.

The idea in the end is simple. Take the multi-particle wave function of the ground state and compute it's square. The wave function is given by
$$
\psi_F = \prod_{i<j}(z_i-z_j) \exp(-\sum z\bar z/2)
$$
Now, we can  calculate a probability density in the phase space of the $N$ particles 
by squaring the wave function:
\begin{equation}
p(\vec x_i) = \prod_{i<j}|z_i-z_j|^2\exp(-\sum z \bar z) = 
\exp(-\sum z\bar z +2\log(|z_i-z_j|))
\end{equation}
The idea now is to go to a coarse-grained point of view of the system. From the form of the function $p$, the square of the wave function can be interpreted as an ensemble of $N$  particles in two dimensions  at positions $x_i$ in the presence of a quadratic potential with pairwise repulsive interactions
at some finite temperature. What we are computing is the classical Boltzman factor for the canonical ensemble of the particles $exp(-\beta H)$. This idea follows exactly the macroscopic description of the quantum hall effect droplets by Laughlin \cite{Laughlin}. See also \cite{QHE}.

 Because this is a statistical mechanics problem with a lot of identical particles, we can hope that the probability distribution $p$ peaks at some configuration which captures the essential coarse-grained properties of the system, which includes the density 
distribution of particles in the geometry, etc. This type of formulation in terms of statistical mechanics assumes that the details of all the particles individually are too complicated
to understand. Instead, we should look for other observables, like the number of particles in some coarse grained volume of phase space. This is what leads to a description in terms of densities. The coarse graining should be such that the volumes considered are quite larger than the  area that a particular particle would occupy in phase space given by the uncertainty relation.

To describe the ground state, there is a preferred density distribution to consider (this is a preferred configuration of our statistical mechanical system).
The idea is that this preferred configuration can calculate arbitrarily well 
all coarse grained observables. These are the observables that change extremely slowly when we vary
the positions of all the particles just a little bit. In essence, they don't depend on microscopic details.

If we think of this statistical system, we want to find the equilibrium configuration of the statistical system which minimizes the energy (including the repulsive interactions). In this statistical framework, we hope that in the end the fermions will form a reasonably shaped object with a continuous density profile $\rho(\vec x)$, so that
sums appearing in $\beta H$ can be approximated by integrals that depend on $\rho$, plus a constraint for the total number of particles to be equal to $N$ which is a very large number. 
This is given by the following
\begin{eqnarray}
\sum_i z_i\bar z_i  &\to & \int d^2 x \rho(x) \vec x^2\\
\sum_{i<j} 2\log(|z_i-z_j|)&\to & \iint d^2 x d^2y \rho(\vec x)\rho(\vec y) \log(|\vec x-\vec y|) \\
N &=& \int d^2 x \rho(x)
\end{eqnarray}
The quantity $\rho(x)$ is constrained to be positive. We can try to remove this constraint by 
writing $\rho$ as a square if we want to. 

The idea now is to write a variational principle for $\beta H$ to find the most likely configuration of particles (the one with least energy will be sharply peaked, because of the large number of degrees of freedom), by varying $H$ with 
respect to $\rho$ and setting that variation to 
zero. This will produce the following integral equation
\begin{equation}
\vec x^2 + C= 2 \int d^2 y \rho(y) \log(|\vec x-\vec y|) \label{eq:variat1}
\end{equation}
which is valid only for those $x$ where $\rho(x) \neq 0$, as where $\rho(x)$ vanishes the variation is constrained. In the above, $C$ is a Lagrange multiplier enforcing the constraint on the number of fermions. Now we can use the fact that we are in two dimensions and that
$\log(|\vec x-\vec y|$ is the Green's function for the Laplace operator. The problem can be interpreted as a Coulomb gas in two dimensions in a particular background electric field produced by some constant density background charge.

 From here, one can take the Laplacian on both sides of \ref {eq:variat1} and use the fact that
$\nabla_x ^2 \log(|\vec x-\vec y| \sim \delta(\vec x-\vec y)$, so that
one finds that in the end $\rho$ is constant on it's domain.  Indeed, the potential $x^2$ is the one we obtain for a uniform charge distribution on the plane, and the charged particles move to a shape that cancels the electric field locally.

Next we can use rotational symmetry of the system to show that  the distribution of particles can be on a circular disk. Finally, one can check for stability of the configuration, to find that the disk is the configuration with minimum energy. The effective electric potential of the saddle point configuration, including the background charge is then 
given by the following figure \ref{fig:potcgas.eps}

\myfig{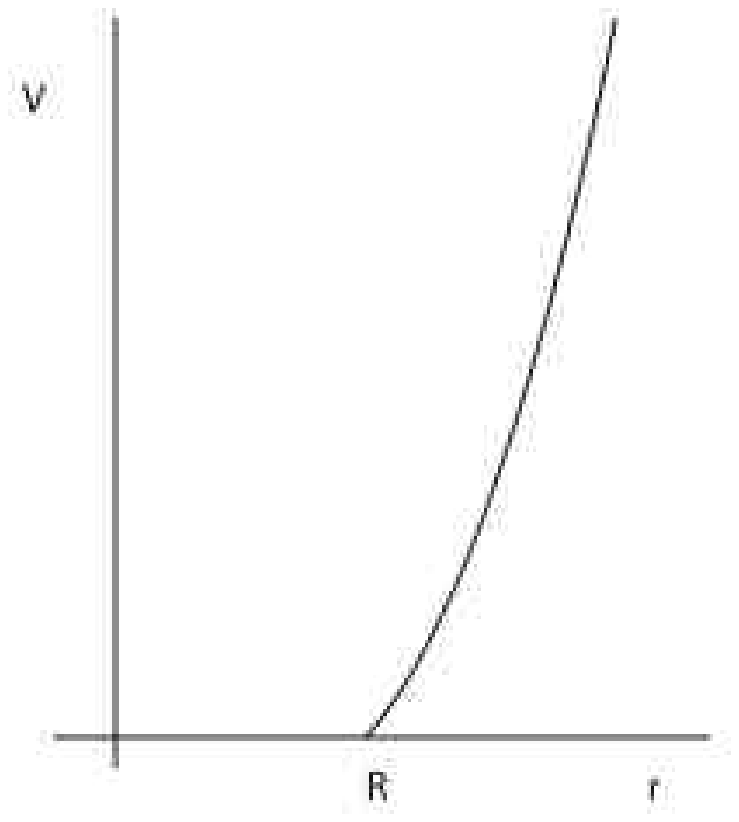}{3}{Effective radial potential $V(r)$ seen by a probe charge. R is the radius of the droplet. The potential is constant inside the droplet.}

The area of the disk in the end is determined by the constraint on the number of eigenvalues, and the constant density is exactly as predicted by quantum mechanics. This method of arriving at the final answer reproduces what we expected from the fact that quantum states occupy finite area, and that Fermi statistics forces them to spread out into a uniform density droplet. 
Notice that we have converted our cartoon of where the particles are to a precise 
mathematical formulation of what we mean by the droplet and it's shape. The droplet is the
saddle point approximation of the square of the wave function of the many-particle system. This is a very important point in this discussion.

The advantage of using this formalism becomes apparent when we try to deform the quantum state by considering a coherent state of the trace oscillators. Coherent states are the way we think about classical linearized fields. The trace oscillators have been identified with the individual gravitons in
the AdS/CFT \cite{AdSCFT2}. The idea now is 
that to produce a classical field configuration for the mode of energy $m$, we take something which is approximately a coherent state, of our approximate oscillator.
This is, we take $a_m^\dagger\sim \tr(Z^m)$ properly normalized, and consider writing the approximate coherent state
\begin{equation}
\ket{\gamma_m} \sim \exp_T(\gamma_m a^\dagger_m)\vac
\end{equation}
We need to be careful not to use the naive coherent states. This is why in the notation above we have introduce a $T$ subindex. A proper exponential of the trace gives rise to a non-normalizable state in the fermion  picture, so we should truncate the expression at some 
high (perturbative) order in $\gamma_m$, where the corresponding term $\gamma_m^T r^m /T!$ 
is a very small number at the radius of the droplet. The idea is that if we were to compare this state to a coherent state in 
a true harmonic oscillator, the error we would be making when we compare the two states is actually very small: the truncated coherent state and the true coherent state have a very high 
overlap.

Given this warning, in the saddle point calculation we will treat it as if it were a proper exponential. Formally one has to take these calculations as a systematic asymptotic series. This is the sense in which the calculation we are doing is a perturbative calculation.
We will play the same game as before: we write the square of this new wave function,  and we can again
interpret it in the form of an ensemble, where the one particle potential is not quadratic anymore, but 
instead we have replaced it by
\begin{equation}
-\beta \vec x^2 \to -\beta \vec x^2 + \gamma_m N_m Z^m+c.c.\label{eq:defpot}
\end{equation}
where $N_m$ is the normalization factor for the mode $m$, and $Z= x+iy$. 
Since $\gamma_m$ is small, we can to a first approximation use the old solution of the droplet with constant density. However, we can not use spherical symmetry any longer to determine the shape of the droplet. Instead, we use the analogy to  charged particles in a Coulomb gas (a 2D plasma) to realize that
we lower the energy if we try to follow the equipotentials of the new effective potential. This is because the local variation in the potential is small compared to the potential that is already there, so the logarithmic repulsion between the eigenvalues will not contribute sizably to small changes of shape.
Since $Z^m \sim r^m exp(im\theta)$ in polar coordinates has $m$ nodes on the circle, the droplet will deform to a shape that is almost circular with a wavy shape on the edge which has $m$ bumps on it \footnote{This is the way in which bosonization is realized in the integer quantum hall effect. A discussion of this bosonization and edge dynamics for this case can be read in \cite{Stoneedge}. }.
 The angle at which the bumps appear depend precisely on the phase of $\gamma_m$. A general coherent state will deform the droplet edge without as many regularities. This deformed shape is the coarse grained description of the wave function at time $t=0$, and then we can let the system evolve in time on it's own. This is exemplified in figure \ref{fig:wavyedge.eps}.

\myfig{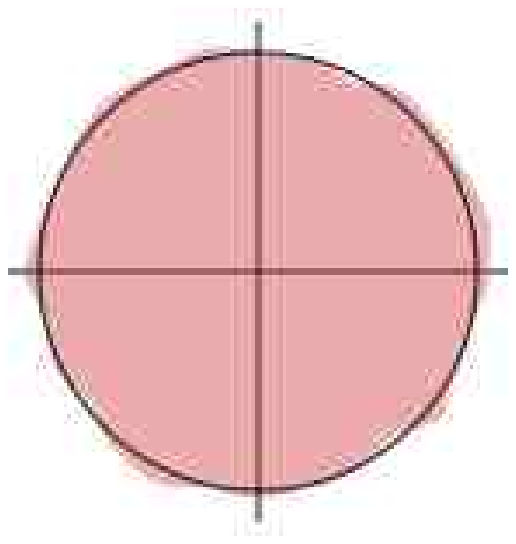}{4}{Distorted edge of a droplet by a coherent state}

More precisely, the effective potential will add a small force to the eigenvalues. This force is strongest near the edge, because it grows with the radius, and of order
given by the derivative of the potential. This is of order 
$m \gamma_m N_m r^{m-1}$. Using the normalization found for the traces before, 
$N_m\sim (m N^m)^{-1/2}$, we find that the force on an eigenvalue on the edge
associated to this coherent state scales like $m^{1/2} N^{-1/2}$. The dependence on $m$ 
follows because the total energy associated to the deformation should be roughly the square of the amplitude. This should also match the energy for individual quanta.
 The fact that the force dependence as a function of $N$ decreases with $N$ is what
guarantees that in the large $N$ limit the deformations can be analyzed using small amplitude waves.

Notice also that we have deformed the potential by adding to it a sum of a holomorphic piece and it's complex conjugate. Thus, when we express the variational principle, we get a modified form of equation \ref{eq:variat1}, namely
\begin{equation}
\vec x^2 + \gamma_m N_m Z^m+c.c. = 2\int d^2y \rho(y) \log(|x-y|)
\end{equation}
Again, when we take the Laplacian on both sides, we notice that the Laplacian of the holomorphic and anti-holomorphic pieces vanish and the droplet still has constant density, but a different shape. This is what makes the fermion droplet incompressible in this coarse grained description.

Notice also that we have a one to one correspondence between coherent states and droplet shapes. Thus to each droplet shape corresponds a unique quantum state determined by the shape of the droplet. These coherent state wave functions are special. They are singled out by the coarse grained dynamics. 

In the AdS/CFT correspondence, it is here that one can compare directly 
to classical linearized supergravity solutions which preserve the requisite amount of supersymmetry. What we see very clearly from this exercise is that coherent states of collective oscillators in the droplet picture associated to the CFT
correspond exactly to shape deformations of the droplet in the statistical ensemble saddle point approximation, while keeping the density constant. These coherent states are represented in supergravity by classical supergravity solutions with the requisite modes turned on. These also match on exactly to shape deformations of the droplet as understood in \cite{LLM}. 

Here we see that spacetime geometry is making an appearance because the eigenvalues (fermions) assemble themselves non-trivially on phase space.  Indeed, the size of the shape that the fermions assemble themselves into is  large in fundamental units. This is why the system can be described in a classical approximation based on geometry. It is the geometry of the droplet in phase space that matters. This is very similar to the $c=1$ matrix model (for a review see \cite{kleb})

 This is happening in a natural coordinate system that arises
directly from the dynamics of our theory. This is to say that 
 the droplet knows also about the harmonic oscillator potential. Without the information associated to the Hamiltonian, the only observable we would have is the area of the droplet, as that is one of the few invariants under canonical transformations. With the hamiltonian flow (understood also as part of the preferred coordinates system) , 
 we can also  describe the angle that the edge of the droplet has with respect to the Hamiltonian vector field, as well as talking about the curvature of the edge. In some sense, the 
 Hamiltonian function can also be considered as a Kahler potential for a flat metric on 
 the quantum plane. All curvatures and geometric invariants are measured with respect to 
this Kahler metric.

To go beyond this perturbative regime at the level of microscopic wave functions in general is hard. This is because  typical wave function terms can not be easily put in the exponential in most cases, and calculating saddle point of many-particle statistical systems is in general hard. 
Coherent states turn out to be fairly tractable for our purposes. It turns out that hole wave functions are very tractable as well. A hole around position $\lambda$ is described by a determinant. This is the wave function of $N+1$ eigenvalues, with the $N+1$th eigenvalue located at $\lambda$ 
removed. This is felt by the other fermions in the Vandermonde determinant. The terms that depend on $\lambda$ are given by  the product 
\begin{equation}
\prod_i(z_i-\lambda) = \det(Z-\lambda) = \exp (\tr(\log(Z-\lambda))
\sim \exp (\int \rho(z) \log(z-\lambda))
\end{equation}
This is also a simple nice holomorphic function of $Z$, which is holomorphic away from $z=\lambda$ \footnote{As a side remark, in the $c=1$ matrix model literature, the operators $\tr(\log(Z-\lambda))$ correspond to the so called macroscopic loops, and they also exponentiate to D-branes. See for example \cite{MTV, MMSS} and references therein  }.

Notice that the main effect of introducing a hole in the picture is that we get a repulsive
force from the location of the hole in the statistical description.
 A wave function associated to many holes located at $\lambda$  produce a large hole in the fermion 
description, because it raises the local potential sufficiently. This is shown in  figure 
\ref{fig:hole.eps}.

\myfig{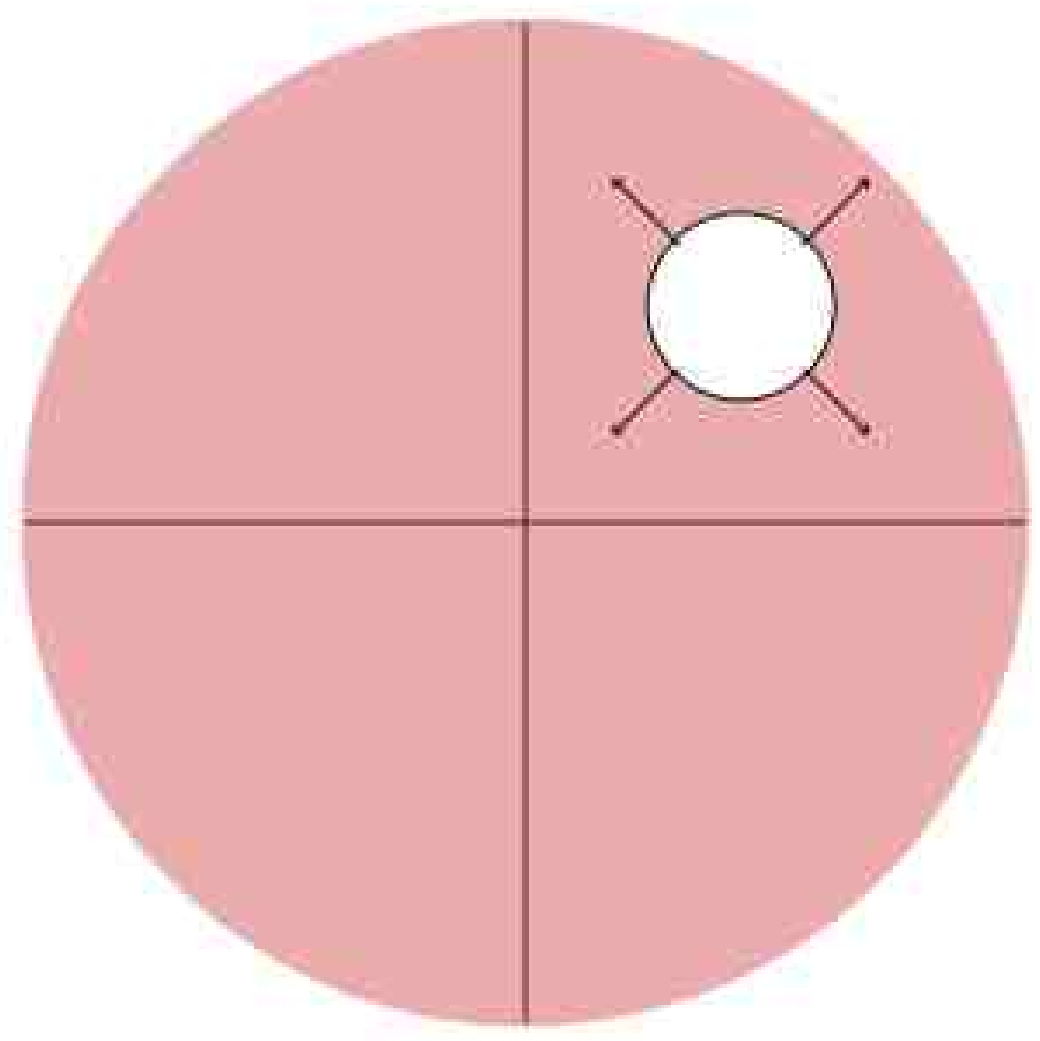}{4}{The hole wave function produces a repulsive force from the hole location}

In general one also has the Vandermonde of the holes to consider, which makes the notion of the shape of stacked holes important, because it accounts for the Fermi statistics of the holes. This can give rise to a new semiclassical edge, which corresponds to a change of topology for the shape of the edge.
This ends up inducing a topology change in the spacetime geometry itself \cite{LLM}. 
If we understand half BPS gravitons as massless particles in the eikonal approximation (moving along null geodesics), there
are now two disconnected sets of null geodesics in the geometry which corresponds to that classical motion. In quantum mechanics this implies that there is a potential barrier between 
them, and in the semiclassical limit we get two sets of excitations: one for each edge.

This description of holes in terms of determinants also produces the usual giant graviton operators \cite{BBNS} associated to D-branes on $S^5$ as a 
power series in $\lambda$. The standard giant graviton wave functions obtained that way are delocalized on the angular direction of the circular droplet in the quantum plane: they are shaped like rings. These are the ones that correspond to energy eigenstates of the Hamiltonian,
and therefore have extra symmetry.

The expectation is that in general for other droplet configurations we can use the coarse grained dynamics to describe the salient features of the distribution of particles, and that this is the essence of what semiclassical gravity is capturing: the leading asymptotic expansion about a coarse grained saddle point configuration, which in our case depends on the wave function of the CFT dual in a very precise way. 

It is also important to notice that in this coarse grained description we have given, configurations of
droplets of density one can arise from very few pure states in the quantum system. This is because we need to pack the Fermions as much as is 
possible in phase space, and this leaves no degrees of freedom left over that could account for various micro-states having the same coarse grained description. This suggests a reason why some half-BPS geometries have singularities. These singular geometries correspond to 
regions of positive  density bellow the critical value \cite{LLM} (a more recent discussion can be found in \cite{MiL}, where the structure of the singularities is clarified), like the half-BPS superstars \cite{Metal}.
In the coarse grained description we have given above, one can find a lot of wave functions which differ 
by minute variations of the local density, which is allowed because the packing of fermions is not tight.  These variations nevertheless can give rise to orthogonal states, so there are many pure states which end up corresponding to the same saddle point:
the coarse grained description would not detect these variations. This suggests that 
the curvature singularities in this case are related to entropy. It is also possible that some special micro-states for these packings might have a nice large $N$ description (for example one could consider the wave function of a fractional quantum hall effect), and could be associated to a stringy geometry regime of quantum gravity.

Now, returning to our results, we need to interpret the procedure we have outlined.
In some sense we can do it by writing a master formula that is supposed to give us our intuition along the following lines:
\begin{equation}
|\psi|^2 \sim Z_{gravity}\label{eq:psi2z}
\end{equation}
and coarse-grained observables as a dictionary of the following sort
\begin{equation}
\int \psi^* \CO \psi \sim \langle \tilde \CO \rangle_Z
\end{equation}
where $\tilde \CO$ can be related to an extensive quantity.
The left hand side is the square of the microscopic wave function of the Universe (in terms of the dual CFT wave function), which exists in a minisuperspace approximation. This can be considered as a minisuperspace wave function a la Hartle-Hawking \cite{HH}, except that in
our case we have $\Lambda <0$ and there is a boundary of spacetime where we can define 
Energy functions, etc. 
The right hand side is a minisuperspace  partition function. This partition function determines the dominant geometry of spacetime by taking a saddle point approximation, and in some sense it is here that gravity is more manifest. In general, one can also try to interpret it as a sum over topologies of spacetime. Indeed, any shape of the droplet will give some number on both the left and the right hand side of \ref{eq:psi2z}. This is a probability that the universe has the appropriate geometry/topology. Since we can superpose geometries on the left by taking some general wave function, we get some
sum of terms on the right hand side that one can try to interpret in terms of tunneling probabilities and persistence amplitudes. In some sense $Z_{gravity}$ as defined above is 
very close to the idea of a Euclidean partition function for gravity. This is in part because $Z_{gravity}$ is a partition function which only gives real numbers associated to 
probabilities, which is also natural in Euclidean field theory, as realizing statistical mechanical systems.
This interpretation is reminiscent of recent work in topological string theory for counting 
 entropy of BPS black holes in four dimensions \cite{BPStop}, and a relation between the topological string partition function interpreted as a wave function, whose square is related to some gravitational partition function. This connection is beyond the scope of the present paper. It certainly deserves to be explored further.

The second thing we would want to understand now, is how having an emergent geometry, the perturbations about it seem local. Again, we need to look at our coherent states and include some more information about the dynamics. In particular, we have not said anything about time dependence of the droplet configurations. This is in some sense trivial. The droplet rotates without changing it's shape about the origin at constant angular frequency. This is just the classical motion of a free harmonic oscillator, extended to all the particles. What is very interesting, is that small perturbations of the ground state geometry move at constant speed at the edge: the angular velocity becomes a geometric velocity because of the extended shape of the fermion droplet . This is the speed of sound of the edge waves. One should understand this speed of sound as a hydrodynamic property of the system. This is also coarse grained, in that we are measuring distances along the edge of the droplet, which is a well defined notion only in the coarse grained approximation.

The advantage of having this type of hydrodynamic description, is that we can also try to do the collective coordinate quantization of the system, e. g., the quantization of the small fluctuations around the ground state. In the condensed matter literature this is how the bosonization of the fermion system is understood. The bosonized description is exact only in the thermodynamic limit. This is the relevant limit for our considerations as well, as well as to systematically go away from it. This collective coordinate quantization has been performed in the supergravity limit in the works \cite{Collect}.

If we associate a geometric radius of order one to the edge, this geometric length scale is identified as the radius of the AdS geometry. This is from reading the size of the droplet shape from a comparison with the corresponding supergravity solution. The speed of sound should be identified with the speed of light in the $AdS_5\times S^5$ geometry. 
At the level of half-BPS solutions this is the most we can really hope for.

Now we also need to recall how this geometry is embedded in spacetime to understand what we are describing. Outside the droplet, the 
coordinate system in gravity is associated to the AdS geometry. Inside the droplet, the information relates more to the $S^5$\cite{LLM}. At the edge, we focus on  a particular set of null geodesics
of $AdS_5\times S^5$. This is the only place where we can match the local excitations we found in our droplet to local excitations in supergravity. Indeed, we need to work in the approximation where massless particles go along null trajectories (the eikonal approximation). The null  geodesics where particles saturate the BPS bound with respect to the supersymmetries we have chosen is almost unique.
All these geodesics go along a particular diameter of the $S^5$. This is a circle, which is identified with  the edge of the droplet. Thus, if we deform the edge of the droplet, it is something related to what happens to the shape of the $S^5$ along this diameter.

Afterwards, the rest of the ten dimensional geometry is reconstructed from the droplet configuration \cite{LLM}.  We are forced to match  the waves we are constructing 
with gravitational waves with some particular profile in the transverse directions.

Also, to send information from one place on the supergravity solution to another, using the modes available to us by considering only half BPS excitations,  two static observers should be situated along this special diameter. Thus, both of them are at the origin in $AdS$, and at different positions on the $S^5$. If the first observer wants to send a signal to the second one, he must have some finite energy which he uses to excite the Fermi sea. The other observer sees the perturbation arrive to him causally in spacetime.  From the microscopic point of view, we need to calculate the group velocity of the collective coordinates describing the coarse grained picture of microscopic dynamics. This is of course constant, and it is associated to the speed of sound in the droplet picture. The corresponding spacetime concept is that the information arrives at the speed of light, which is the speed at which gravitational waves propagate.

The last thing we need to consider in this section is when does the coarse grained picture start to fail.
This is easy to understand as well. There are two ways in which we can expect to see the failure 
of the description. The first case, is when we consider sending waves with very high wave number, so that the wave is sensitive to the individual microscopic positions of the particles on the edge. There are of order $\sqrt N$ particles on the edge (if each one occupies a small round circle on the plane whose are is of order $\hbar$). Therefore, when momenta of individual gravitons gets to be of order of $\sqrt N$ 
we should expect to start seeing something that tells us that the droplet is a coarse-grained description of the dynamics. If one looks at planar diagrams for single traces, this is the place where non-planarities become of order $1$ \cite{J2N,BN, toyads}. The non-planar diagrams are enhanced by the fact that these states are described by very large quantum numbers.

Similarly, we could consider adding an eigenvalue sufficiently separated from the edge, so that a circle of area $\hbar$ about it does not intersect the big fermion droplet. The energy associated to this state
is of order $\sqrt N$ larger than the topmost eigenvalue of the droplet. This is because the energy at radius $R$ is of order $R^2$. At radius ${\sqrt N}+1$ the energy is of order $\sqrt N$ larger than the energy at the edge of the droplet. We could also consider a hole and get the same estimate.\footnote{ Eigenvalues (fermions) and holes in the string theory are interpreted as D-branes \cite{BBNS, CJR, toyads}. In particular, these D-branes are the giant gravitons \cite{gg}. There is by now overwhelming evidence for this identification \cite{giants, LLM}.}

 It should not be surprising that both calculations of  energies coincide. In a perturbative approach, non-perturbative effects become relevant when the effective coupling constant becomes of order one.
The effective coupling constant in this case is related to $1/N$ and the energy of a particle:
the $1/N$ expansion is in $J^2/N$, where $J$ is the angular momentum of the graviton.
 D-branes are the leading non-perturbative effect in weakly coupled string theory and they can be BPS (supersymmetric).
 
We could at this point try to claim success for our program. We have explained locality in a mini-mini-superspace approximation. We have followed a route where we have been able to identify the droplet shape degrees of freedom as a coarse grained description of the microscopic dynamics, and we have a heuristic description of gravity in the half BPS sector.
The description of the ground state, or of the LLM geometries is done in terms of droplets of density one.
This leads us to hydrodynamical concepts, like local densities of particles and shapes,  and to consider transport phenomena in non-relativistic systems (the fermions that make the droplet are non-relativistic).  The picture is very compelling and already has shown us a lot of lessons to be applied elsewhere.

There are some other results available in the literature related to these half BPS geometries
that we have not mentioned yet. For example, in \cite{CKS}, the authors found that droplet configurations with overdense fermions or with negative densities (overdense holes) are unphysical as they violate the Pauli exclusion principle. These were correlated with half BPS geometries with closed timelike curves. This gives us some relation between chronology protection and the microscopic dynamics of quantum gravity.
Also, in \cite{HS} the leading topological transition metric was identified and studied, and it was suggested that there is some topological twisting of the gauge theory which captures all of the half BPS sector dynamics and relates it to the non-critical $c=1$ matrix model. On a more speculative note, \cite{BJS} have proposed a different notion of coarse graining of geometries 
than what we have done so far, based on information theoretic notions, 
that might explain some aspects of black hole entropy and 
microcanonical counting of states. These results are clearly related to the discussion we have done in this paper, and offer another point of view to attack slightly different problems in quantum gravity than we have, mostly related to quantum information and the 
thermodynamics of black holes. Some of their unpublished results overlap with the discussion
above for half BPS states.

In \cite{TY} a systematic approach to
expand the supergravity solutions in the LLM coordinate systems around a Penrose limit was initiated, a step which might be important to further our understanding of string theory in supergravity backgrounds. Finally, the half BPS solutions with $AdS_5\times {\mathbb{RP}}^5$
and their dual formulation have been studied recently in \cite{Mukhi}. 

Regarding the relation to the quantum hall effect and the CFT matrix model, in 
\cite{IMc} a proposal was made for a string dual to the large $N$ harmonic oscillator quantum mechanical system. Also, in \cite{Ghosi} a more systematic analysis of the relation between 
the quantum hall droplet picture and supergravity was started, particularly to understand other pictures of the quantum hall dynamics and their relation to supergravity and in \cite{TT} a more detailed analysis of the relation between the complex matrix model and free fermions was studied. Also, in \cite{DJR}, a first attempt has been made to go beyond a one matrix model
dynamics for the AdS/CFT.

What is missing from the above picture? We seem to have good control of supergravity for the half BPS sector, including non-perturbative topology changes. However, the AdS/CFT correspondence is not just supergravity. It is string theory, and
we have only supergravity modes when we consider the half BPS states.
We need to show also
how the massive strings appear, and how do they understand the target space geometry. We have not explained the dynamical origin of the string scale either. Also, the dynamics has so much supersymmetry, that anything that has the requisite quantum number at the free field level, is protected. This means that any information about 
the SYM coupling constant has been lost.

To compensate for all of these things, we need to go beyond $1/2$ BPS states. Thus, we will study the system with $1/4$ BPS supersymmetry systematically. For at least part of the way, we can study also the system with 
$1/8$ BPS supersymmetry as well. This will be our minisuperspace arena in the rest of the paper. It is also at this level that 
going from the free field limit to the interacting theory becomes non-trivial. Some quantum states that  saturate the $1/4$ and $1/8$ BPS bound in the free field theory limit get lifted
and can become strings. This is where the string scale is going to start showing up. We will also try to describe these objects in an analog of the droplet picture
to get some more intuition about them. The tools that we will use for this generalization are straightforward extrapolations of the techniques studied in this paper so far. Conceptually we will be copying the procedure as much as possible.
Some technical details are harder to understand, and lead to novel ways to see how classical spacetime geometry emerges.

\section{ The chiral ring: BPS states}\label{sec:chring}

The ${\cal N}= 4 $ SYM theory  with gauge group $SU(N)$ 
can be considered as a special case of an ${\cal  N}=1$ SYM 
theory with three adjoint superfields $X,Y,Z$ and a superpotential given by 
$\tr( X[Y,Z])$. As an ${\cal N}=1 $ theory, we can ask what the 
allowed vacuum configurations of the theory are, this is, what the moduli space of vacua 
of the theory is (this is for the theory on flat space). This question is usually answered by stating that the moduli space of vacua is the set  of all representations of the chiral ring. We will get back to this type of answer later on. However, first we will answer the problem classically: what are the classically supersymmetric vacua of the theory?

The classical moduli space of vacua  is given by solutions of the F-term constraints $[X,Y]= [Y,Z]= [Z,X]=0$ 
which tells us that the moduli space is parametrized by a set a three commuting matrices
(which are traceless dues to the fact that they are in the adjoint of $SU(N)$). One also needs to worry about the D-terms, this makes it possible to diagonalize the matrices simultaneously by using a unitary transformation.

We can choose the gauge where all the three matrices are diagonal, and then parametrize the solutions to the vacuum by the eigenvalues of the matrices. To each diagonal element of the triplet of matrices we can assign a point in $\BC^3$.  
However, once this is done, there is still  an unbroken symmetry which permutes the eigenvalues, so that the moduli space is a symmetric product space $(\BC^3)^N/S_N$ for the $U(N)$ matrices. 
To get the result for $SU(N)$, we can mod out by global translations (e.g. we can choose an origin about which the dipole moment of the eigenvalue distributions vanish).

This characterization of the vacuum in terms of eigenvalues is one way to parametrize the vacuum configurations. This requires us choosing a gauge and the results depend on how this is done. However we can also classify the vacua in terms of gauge invariant operators. This is the characterization which is given in terms of the chiral ring.

The chiral ring is the cohomology of the superspace derivative $\bar D $. For our purposes this will be the set of all local holomorphic gauge invariant operators
built by polynomials of the $X,Y,Z$, modulo the F-term equations (these are total derivatives under the $\bar D$ operation). These are generated by traces of the form \begin{equation}
{\cal O}_{n_1,n_2,n_3}= \tr(X^{n_1} Y^{n_2} Z^{n_3})
\end{equation}
In the eigenvalue basis this is the set of all multipole moments of the eigenvalue distributions.
This is automatically invariant under the residual $S_N$ transformations. The traces are not
algebraically independent. One can generically choose a basis of $3N-3$ such traces (the dimension of the moduli space) which one can use to parametrize a set of local coordinates in the moduli space, and all others can be obtained from these in a small neighborhood of a regular point.

The ${\cal N}= 4$ theory is a conformal theory at the quantum level, and the generic vacuum breaks conformal invariance spontaneously.
However, there is an origin in moduli space where all eigenvalues vanish, and it is here that 
the theory has full conformal invariance. The operators that can acquire a vev in the chiral 
ring turn out to be protected operators in the conformal field theory. These satisfy the BPS bound $\Delta = (3/2)R= J_1+J_2+J_3$  for a particular $U(1)_R$ charge, which is embedded diagonally in the $SO(6)$ R-charge symmetry group of the ${\cal N}=4 $ SYM theory.
This is given as follows in the fundamental representation of $SO(6)$,
\begin{equation}
3/2 R \sim \begin{pmatrix}J_1& 0 &0\\
0 & J_2&0\\
0&0& J_3\end{pmatrix}\sim\begin{pmatrix}\sigma_1& 0 &0\\
0 & \sigma_1&0\\
0&0& \sigma_1\end{pmatrix}
\end{equation}
where $\sigma_1$ is one of the Pauli matrices. The generators $J_1, J_2, J_3$ are
the Cartan basis of $SO(6)$.

The proper statement that one should make for operators is not exactly that the cohomology of $\bar D$ is protected, 
but that there is always an operator with the given quantum numbers of an element in the chiral ring which is protected.
Remember that the chiral ring can not distinguish between $\tr (X^2 Y XZ)$ and 
$\tr (X^3 YZ)$, because they differ by an F-term,  but that these are different operators in the ${\cal N} =4$ SYM theory. Also, as argued above, multi-traces are allowed because they get a vev on the moduli space of vacua as well.

There are additional generators of the chiral ring. These involve the chiral $W_\alpha$  field 
strength from the gauge fields, whose lowest component is a gaugino superfield. A good discussion of this can be found in  \cite{CDSW}. 
For example, one can have single trace operators 
\begin{equation}
\tr(W_\alpha X^k Y^m Z^n)\quad \tr(W_\alpha W^\alpha X^kY^mZ^n)  
\end{equation}
And the second set of operators
acquire vevs in generic ${\cal N}=1$ gauge theories because they are Lorentz scalars. These give rise to gaugino 
condensates and generically can be related to (partial) confinement.

Using the operator-state correspondence, we can turn these protected operators to protected states for the ${\cal N}=4$ SYM theory on $S^3$. These operators can be the highest weight state of $1/2,1/4$ or $1/8$ supersymmetric states depending on the quantum numbers of the state with respect to the $SU(3)$ R-symmetry group,  which commutes with our choice of $R$ charge for the ${\cal N}=1$ projection.

They are half BPS if they saturate the BPS bound and  $J_2=J_3=0$, quarter BPS if $J_3=0$, 
and $1/8$ BPS otherwise.

 The elements of the chiral ring are all highest 
weight states with respect to the ${\cal N}= 1$ superconformal group, but not necessarily with respect to the ${\cal N}=4$ superconformal algebra (some are related by $SU(3)$ rotations to others, so they belong to the same multiplet). 

It turns out that  every $1/8$ BPS state in the ${\cal N}=4 $ SYM theory can be obtained as
descendants of these operators. The case of quarter BPS operators has been studied extensively in \cite{quarterBPS}. 
If we understand these $1/8$ BPS operators, we have understood the counting of $1/8$ BPS states in the theory.
One realizes quickly, along the lines of reasoning in \cite{toyads} that all of these operators are built only out of the 
$s$-wave of three complex scalars on $S^3$, plus the possibility of a single partial wave for fermions (these have two different polarizations of spin up or down).

This is because the complex scalar have dimension one and 
$R$-charge $2/3$ ( J-charge one), while 
some of the spinors have dimension $3/2$ and R-charge one (J-charge equal to $1/2+ 1/2+1/2$.

 Other partial waves on $S^3$  appear as the descendants of the operator (they add to $\Delta$ but not to $R$), as they are related to derivatives of the fields under the operator state correspondence. These don't contribute to the essential dynamics of $1/8$ BPS 
states, and therefore one should integrate them out. Thereby one is reducing the problem to the study of three matrices at a point, one each for $X,Y,Z$, plus two fermionic matrices, associated to the $W_\alpha$.

All the true quarter BPS operators are
multi-trace operators:
if we restrict ourselves to  the single trace operators that can show up,  we find that they  are an element of a  $1/2$ BPS multiplet (if we insist on purely bosonic states of the form $\tr(X^{n_1} Y^{n_2} Z^{n_3})$, they are obtained by $SU(3)$ rotations of $\tr(Z^k)$ for $k= n_1+n_2+n_3$). This is due to the 
fact that other possible operators with different orders in the letters have a non-trivial one loop planar anomalous dimension and are not
protected. This can be shown explicitly from the $SU(3|2)$ spin chain model for the set of states in question \cite{Beietal}.  Perhaps surprisingly, once we solve the planar problem, the non-planar problem of calculating the anomalous dimension does not lift states from the list we have considered up to this point.

For the AdS dual geometry, this means that all of these $1/8$ BPS states are obtained from collective excitations of supergravity fields, so they might be argued to correspond to supergravity solutions with asymptotic $AdS_5\times S^5$ geometry which preserve $1/8$ of the supersymmetry.  Of course, it would be very interesting to set up this correspondence with the full $1/8$ BPS solutions of supergravity. However, the supergravity story is not available yet. This work should be considered in some sense as only one half of the story:  the CFT droplet picture for solutions with less supersymmetry. In the half BPS case,  this is the boundary condition
data for a partial differential equation which reconstructs the ten dimensional geometry. For our purposes we will assume that something similar happens in this less supersymmetric case. What we will be interested in, is in developing the same tools that were used for the SYM description of half BPS states in this more complicated setting  and we will try to argue how classical physics on the $S^5$ becomes local.

With all the considerations we have made above, we will now study the (gauged) dynamics of the
$s$-waves of the scalar fields of ${\cal N}= 4$  SYM compactified on $S^3$, which we have argued is the correct dynamics to study, and in the end we will reduce these to the dynamics of three real variables plus their conjugate momenta in phase space. Afterwards we would need to add the two fermionic matrix variables as well. This is a consistent truncation of ${\cal N}=4 $ SYM on the sphere at the classical level. One is only turning modes that respect an $SO(4) $ spherical symmetry for the scalars, and for the fermions one is respecting an $SU(2)$ symmetry on the $S^3$.
At the level of spin chains for the ${\cal N}=4 $ SYM, three scalars with two fermions  is claimed to be a closed subsector of operators  with $SU(3|2)$ symmetry.

We will first look at the classical system (ignoring the fermions). We need to consider the dynamics of  six real scalar matrix variables (the constant modes of all scalar fields on the $S^3$). From their lagrangian we obtain
\begin{equation}
L =\frac 12\tr\left( \sum_i (D_t\phi_i^2 - \phi_i^2) -\frac 12\sum_{i,j} [\phi_i,\phi_j]^2 \right)
\end{equation}
In the above we have made the reduction to only the s-wave modes on the $S^3$. We also need to keep the time component of the covariant derivative, which does not contribute physical degrees of freedom, but provides the Gauss' constraint that the states need to satisfy. This is what reduces the problem to the study of multi-traces of the variables.

In complex notation, this is
\begin{equation}
L =\tr\left( \sum (D_t\phi D_t\bar \phi - \phi\bar\phi) -\frac 12\sum [\phi,\phi][\bar\phi,\bar\phi]  -\frac 12 (\sum [\phi,\bar\phi])^2\right)
\end{equation}
or in Hamiltonian form (ignoring  the gauge field)
\begin{equation}
H =\tr\left( \sum ( p_\phi p_{\bar \phi} + \phi\bar\phi) + \frac 12\sum [\phi,\phi][\bar\phi,\bar\phi]  + \frac 12 (\sum [\phi,\bar\phi])^2\right)
\end{equation}
where we run over three complex fields $\phi\sim X,Y,Z$. The decomposition mirrors  the F-terms and D-terms of the ${\cal N}=4 $ SYM theory.

The $U(1)_R$ charge is given by the generator
\begin{equation}
R \sim (-i) \tr(Z\dot {\bar Z}-\bar Z \dot Z)+(Z\leftrightarrow Y) + (Z\leftrightarrow X)  
 =  \tr (Zp_Z - \bar Z p_{\bar Z})+\dots
\end{equation}
so that $Z$ has charge $1$ and $\bar Z$ has charge $(-1)$, etc. 
From here, let us try to understand how one might saturate the classical BPS bound 
$\Delta\geq 3/2R$. Here $\Delta$ is the dimension of the operator (energy for the states), while $R$ is the R-charge of the corresponding states.
Clearly $\Delta$ is positive, since it is a sum of squares.
If we ignore the commutator terms (arguing by a perturbative reasoning), we get a sum of harmonic oscillators with $\omega=1$
, and we can decompose the modes as follows
\begin{equation}
Z= Z_+ \exp(i t) + Z_-\exp(-i t)
\end{equation}
A similar decomposition holds for $X,Y$, while $\bar Z, \bar X, \bar Y$ will be described by the complex conjugate of the $Z, X, Y$ variables. It is easy to show that the energy will be proportional to 
\begin{equation}
\tr(|Z_-|^2+|Z_+|^2)+(Z\leftrightarrow Y) + (Z\leftrightarrow X)
\end{equation}
while the R-charge will be proportional to 
\begin{equation}
\tr(|Z_-|^2-|Z_+|^2)+(Z\leftrightarrow Y) + (Z\leftrightarrow X)
\end{equation}
so that we need $Z_+=X_+=Y_+=0$ if we want $H=J$. 
If we now include the commutator terms as a perturbation of the dynamics, the
energy increases if the commutators between $X_-, Y_-, Z_-$ don't vanish, while the value of $J$ stays invariant. Thus to saturate the classical BPS bound, we need to require that $X_-, Y_-, Z_-$ commute. \footnote{Presumably this analysis would simplify if one used the corresponding supersymmetry variations and set them to zero, and the argument would not be perturbative. However the author of the present paper is not aware that such a formalism has  been developed in the literature for this compactification of ${\cal N}=4 $SYM }
This observation is crucial for the development of this paper.

In essence, the BPS dynamics (for the bosons) is reduced to the study of a matrix model of three commuting holomorphic matrices. Again, the gauss' constraint will let us 
diagonalize them all simultaneously by a Unitary transformation, so we have in the end a system with 3 normal commuting matrices.
This should not be surprising, because it matches the intuition from the chiral ring, that 
the order of matrices in the traces does not matter, hence they should be thought of as commuting matrices.
The reduction from six scalars down to three is because we are setting $X_+, Y_+, Z_+$ exactly to zero. This should be interpreted as Hamiltonian reduction for these modes.

Commuting matrices also parametrize the vacuum configuration of the BFSS matrix 
model \cite{BFSS}. Indeed, in that case the moduli space of vacua was $N$ copies of the moduli space for a single eigenvalue, and this was interpreted as the space of the $11$-dimensional geometry. Individual eigenvalues were interpreted as both $D0$ branes, and as 
partons of a graviton, if a lot of them where moving together with a normalized wave 
packet.

In our case it is also easy to show that if we make a commuting ansatz for all 
the matrices, we can solve exactly the equations of motion of the classical system, which reduce to
$N$ three dimensional harmonic oscillators (this is for $U(N)$, while for $SU(N)$ we need to remove the trace modes, and we get $(N-1)$ three dimensional harmonic oscillators). 
What the eigenvalues do should be related to the geometry of spacetime by analogy with 
the BFSS matrix model. We will pick this theme later on.

Adding the fermions to obtain all the $1/8$ BPS states at this point is straightforward. We will have a $(3|2)$ dimensional harmonic oscillator per eigenvalue.
This will happen because the off-diagonal fermions will not saturate the BPS bound, as they will be more massive than the ones on the diagonal. This also matches the idea that in the chiral ring we can have only up to two distinct $W$ inside each trace, and that their order does not matter.

One can now give a naive model of all the $1/8$ BPS states: they are in one to one 
correspondence with the set of $N$ free (super)-bosons in a three dimensional 
harmonic oscillator (one still needs to mod out by global translations if one wants the result for $SU(N)$). The three angular momenta associated to each eigenvalue are related to the three directions for the oscillators, and they add up to 
give a Cartan basis for $SU(3)\times U(1)$. The naive group of symmetries is $SO(3)\times U(1)$, but one needs to consider the variables as complex matrices with first order dynamics instead. This is the idea that string theory geometry might make more sense in the phase space of the (super)-boson dynamical system (as in the $c=1$ matrix model \cite{Pol}), so one thinks of the symmetries of phase space which keep the Hamiltonian and the symplectic form (complex structure) invariant. 

For each eigenvalue, we  will have three quantum numbers that characterize the state 
$(n^i_1, n^i_2, n^i_3)$, one for each harmonic oscillator, but we also need to remember the bosonic nature of the 
eigenvalues, so that permutations of triples give equivalent states. We can take care of this 
by  defining a total order for the energies $(n_1, n_2, n_3)>(n'_1, n'_2, n'_3)$ if either $n_1>n_1'$ or
$n_1 = n_1'$ and $n_2>n_2'$ or $n_1 = n_1'$ and $n_2=n_2'$ and $n_3>n_3'$.
Thus we can order the states in descending order, where
$\vec n ^1\geq \vec n^2 \geq, \dots \vec n^N$.
If we use just the $n_1$ quantum numbers for the bosons, we can again get a Young tableaux.
The other labels will decorate the rows with extra numbers. However, there does not seem to be a nice uniform way to collect all eigenvalue information systematically this way.

Another possible counting is to associate to each triple the weight $n_1+n_2+n_3$ and order them in this way. We then use the inequalities $(n_1+n_2)\geq (n_1'+n_2')$ and $n_1\geq n_1'$ to decide how to order the states. This will again give us a Young tableaux with each boson weighed by the energy it carries. We can then paint each  row in the tableaux in three colored stripes, with $n_1$ boxes in yellow, $n_2$ boxes in blue and $n_3$ boxes in red, but again, no nice coloring pattern seems to appear.

In any case, it is easy to see that the counting of states of the bosons, and the counting of states of the traces gives the same results for low numbers of boxes. For large traces, identities will reduce the number of algebraically independent traces at order $N$. Our claim is that the boson eigenvalues we have described above will count all states correctly, including all
the redundancies (algebraic identities) in the chiral ring.

For small energies compared to $N$, most of the vectors are going to be zero.
We can compare these states to $\prod_i \tr \phi^{\vec n^i}$, where we interpret
\begin{equation}
\tr (\phi^{n_1, n_2, n_3}) \sim \tr(X^{n_1} Y^{n_2} Z^{n_3})
\end{equation}
Again, the ordering is allowed here because all of these operators commute with each 
other.
This result above is for the counting of states for the $U(N)$ theory. 

For the $SU(N)$ theory the analysis is more involved, as one needs to mod out the translation mode correctly. But this just removes one triple of oscillators (associated to the center of mass), and this is just correlated to the traces $\tr (X)$, $\tr(Y)$, $\tr(Z)$ which vanish for $SU(N)$ and don't appear in the list of operators in the chiral ring. A detailed version of this procedure has been carried out in the half BPS case \cite{Demel} 

This naive  model should be interpreted in the same way as the naive model of bosons discussed in  the previous section for half BPS states. It counts correctly the energy and R-charge of the states, as well as the degeneracies, but it has the wrong 
operator algebra. From this point of view, supergravity will probably not realize this model, but something 
different instead. However, already here we can count states and match our counting with other results available in the literature. To my knowledge, the counting of $1/4$ and $1/8$ BPS operators has been carried at low orders in \cite{quarterBPS}. Their counting matches the results presented here, where it is easy to generate an ordered list of operators.

This information is already strong enough to argue about statistical properties of these BPS states.
We can now try to count if there are $1/4$ or $1/8$ BPS black holes of large finite area. The typical energy of such a black hole (of radius one in $AdS$ units)  is of order $N^2$. This is from comparing to a thermal state at temperature $T$ of order one in the free field dual CFT. Also, the entropy of such a state should be of order $N^2$, if the black hole is not too different from a neutral black hole. With our description in terms of eigenvalues, at such high energy the statistics of the superbosons don't matter, and we can use Boltzman statistics. We can then calculate how many
states we have at energy $E\sim N^2$. The counting of states is done by calculating the area of a sphere on a $3N$ dimensional space of radius $\sqrt E$. This scales like $E^{3N/2}$, so that the entropy
is of order $S\sim 3N\log(E)<<N^2$. This is too small compared to a macroscopic area of a black hole with finite horizon area in the dual AdS geometry \footnote{Harvey Reall and Radu Roiban have independently arrived at this result }. 
This can serve as a proof that all $1/8$ BPS black holes don't exist. In these cases one can expect that the associated horizon is singular and the would-be  black hole is not black, but that it resolves into smooth microscopic supergravity configurations very close to the singularity, and that the singularity arises from coarse graining of microscopic geometries, similarly to Mathur's program on $AdS_3$ black holes \cite{Mathur} (see also \cite{LMM}) and what happens with generic half BPS supergravity solutions \cite{LLM,MiL}.

Now what we need to do is derive the correct ``quantum droplet" description of the dynamics. The one we have given above  will give the wrong results when we compare to the large $N$ free field theory correlators even for half BPS states. This is a result that we believe we understand well, so the dynamics should be richer.

\section{A quantum matrix model for commuting matrices }\label{sec:matrix}

As we have argued in the past section, to count $1/8$ BPS states we need to consider a 
matrix model of commuting matrices. From the SYM point of view, we have seen that non-trivial commutators contribute to the energy of a state, but not to it's angular momentum, and therefore they take us away from the BPS bound. This is important for finding strings too, as we will see in more detail later on. If we look at the simplest
 operators which are close to being BPS, they correspond to multi-graviton states which are not
 mutually BPS. However, if we choose to keep commutators, the available number of degrees of freedom becomes a lot larger, allowing us to believe that these are the degrees of freedom that contribute to making strings massive with all their possible polarizations. We also expect that in this sector we have objects which are very close to being BPS as well, because many of these operators can be taken in the plane wave limit, where one can engineer the states to have arbitrarily small anomalous dimension \cite{BMN}. For the time being,
 we just want to understand the system given by considering commuting matrices where we have argued that the BPS operators lie. 

We need understand the system in detail, and we have to deal with the fact that 
we need to integrate out the fields that contribute to commutators. Because in the end we are studying BPS configurations, one can argue that integrating these fields together with the rest of the of the SYM multiplet will result in cancellations, so that a careful semiclassical 
argument probably gives the right answer. This is a working assumption. For calculating energies of BPS states, this is correct: the classical result should match the quantum result.
For dynamics, almost certainly not. Otherwise we would expect that a lot of quarter BPS and $1/8$ BPS state observables would be protected by SUSY. This does not seem to be the case.

If we consider that we have a total of 32 supersymmetries, half BPS has the same supersymmetry as the
generic ${\cal N}=4 $ SYM vacua. There are a lot of nonrenormalization theorems in this case ( terms with
up to two   derivatives in the classical low energy effective action are protected). This implies that we should be able to solve a lot of the dynamics exactly. Indeed, 
we have discussed these issues earlier in the paper.

For $1/4$ BPS, the amount of supersymmetry is the same as that of ${\cal N}=2$ SYM in four dimensions. The ${\cal N}=2$ theory is solved by a holomorphic function (the prepotential), and it's holomorphic properties were enough to find the prepotential exactly by Seiberg and Witten \cite{SW}.
The vacuum structure in this case is already non-classical, but the theory is tractable, and one can find the masses of various particles. We expect that in the case of $1/4$ BPS states 
most of what we will describe should be taken with a high degree of confidence.

 For $1/8$ BPS, this is the same as ${\cal N}=1 $ SYM in four dimensions. In the ${\cal N}=1$ theory we expect some amount of non-renormalization which might let us say something about the vacuum of the theory and to calculate the superpotential non-perturbatively. However, in this case there can be a lot of corrections in Kahler terms etc,  that make most calculations intractable. For the $1/8$ BPS states we also need to be careful with setting commutators to zero. This is because we will have fermions. Indeed, the protected operator associated to $\tr(W_{\alpha})^2$ is not that operator as written, but the holomorphic term
in the Lagrangian given by
\begin{equation}
\int d^2 \theta \tr (W^2)+g_{YM}\tr(X[Y,Z])
\end{equation}
This is obtained from the supermultiplet descendants of $\tr(Z^2)$. Indeed, commutators involving three holomorphic fields and bi-fermions mix at second order perturbation theory in planar diagrams at the level of the spin chain models in ${\cal N}=4$ SYM \cite{Beietal}.
The analysis with commuting matrices can still be done in the same way for $1/8$ BPS states
 as for $1/4$ BPS states, but as said above, it's not clear that this will be enough to describe the dynamics precisely. We will not worry about this in this paper. 
 Instead, we will show that the analysis seems to work well for the set of states which don't have the fermions turned on, which was the situation which might require some more care
 to describe the dynamics exactly.

Our assumption of a semiclassical argument is  a place to start the study of the problem. The hope is that
already here some features of the dynamics are apparent and we can give a heuristic description of what is going on, which can later be improved by a more systematic 
expansion. We will see that this seems to be the case, and that we will arrive at 
a very intuitive description of the dynamics which seems to match various aspects of the supergravity dual description.

The way we will setup our semiclassical calculation is to imitate as much as possible the 
calculations done for the half BPS case. To get the leading semiclassical calculation, we just need to include measure factors correctly.
This proceeds by changing variables from a 
generic set of commuting matrices to a set of diagonal matrices, so that the integration measure for the diagonal matrices will be equal to the volume of gauge orbit.
We will later show evidence that this seems to give the correct answer for some questions.

We want to consider a system of either two or three commuting Hermitian matrices, $X, Y, Z$. 
In the supersymmetric context these will be replaced by complex matrices, but these will 
include both the $X$, $p$ coordinates of phase space together, so that it has the same number of degrees of freedom as  three (real) Hermitian commuting matrices and their time derivatives.
The volume of the gauge orbit is easier to calculate if we think only in terms of the wave function of the system in the position basis, so that is the place where we will do our calculation.

We want to write the truncation of ${\cal N}=4$ SYM theory to two or three of the complex scalar fields $X,Y,Z$ which are an s-wave on $S^3$. Our hamiltonian will be associated to the BPS bound $H=\Delta-J$. The set of states with $H=0$ are the set of all BPS states. This is like a Landau degeneracy problem for a lowest Landau level problem in quantum mechanics. 
This is also because a term in the Hamiltonian which couples to angular momentum is of first order in momentum, and can be seen to induce a gauge potential in the Lagrangian. This is very similar to the usual orbital coupling of particles to a magnetic field.
Since $\Delta$ and $R$ commute, we can split the degeneracy problem by calculating the $R$ charges of all the states and organizing them in increasing order.

The associated matrix theory is gauged. The only term of the gauge field that is needed is
the s-wave of the time component of the gauge field, which imposes the Gauss constraint. This is done to ensure that we are dealing with gauge invariant operators in the field theory.
This is similar to the study of the half BPS case \cite{toyads}.

For relevant configurations of commuting matrices, we choose a gauge where they are mutually diagonal. This is a change of variables in quantum mechanics. This change of variables
produces a measure factor in the quantum mechanical problem. This can be associated to integrating out the Fadeev-Popov ghost terms, which give rise to the volume of the gauge orbit of the configuration.

The way we will calculate the volume of the gauge orbit will be 
to choose all matrices to be diagonal and then do an infinitesimal rotation with the broken generators of the gauge 
group, with generators $\theta_{ij}$ for each pair of eigenvalues.

If $X, Y, Z$ are diagonal with eigenvalues $x_i, y_i, z_i$, then the off-diagonal variations will be proportional to 
\begin{eqnarray}
\delta(X_{ij}) &=& (x_i-x_j)\theta_{ij}\\
\delta(Y_{ij}) &=& (y_i-y_j)\theta_{ij}\\
\delta(Z_{ij}) &=& (z_i-z_j)\theta_{ij}
\end{eqnarray}
In principle, there are other off-diagonal components that we can consider when we vary 
  $X_{ij}, Y_{ij}, Z_{ij}$. We can choose these to be orthogonal to the 3-vector given by the differences of eigenvalues of the matrices. If we have various matrices, we will call them collectively $\vec \phi$, with diagonal components $\vec \phi_i$. 
  
  The 
  additional variations are associated to commutators $[\phi,\phi]$, and they don't appear in the case of a single matrix model because there are not enough variables in that case, 
 but they are part of the full $SU(N)$ dynamics of the theory for more matrices.
 
 One can  explicitly check that for the choice of orthogonal vectors to the eigenvalue difference vectors,  these type of fluctuations give rise to non-vanishing commutators, and as we have argued, they should be integrated out. Because of supersymmetry, one can expect that these integrations don't
  change the effective action of the $\vec\phi$ substantially in some regimes mostly because bosons and fermions will cancel each other out.
  
  Now we need to discuss under what conditions this calculation is valid.
   The procedure we have outlined is approximately correct if the 
off-diagonal  modes we are integrating out are heavy. Thus, this is valid in the regime where the 
masses of off-diagonal matrix entries is large. This is like analyzing the Coulomb branch of a field theory in the low energy limit.
The strategy we are following can be compared to the way one solves and ${\cal N}=2$ SYM theory \cite{SW}, this is like expanding around infinity in moduli space and calculating the perturbative correction to the effective action systematically.

Because there is a reduction to commuting matrices, one can now expect that the spacetime geometry will be reconstructed from
what the eigenvalues are doing, a la BFSS\cite{BFSS}. Indeed,  by integrating out  off-diagonal terms one can  recover the gravitational interactions between objects (see for example \cite{BC} for a calculation in various dimensions).

Since in our case we have a field theory in $3+1$ dimensions, and we are studying a
matrix model associated to having everything moving uniformly on the $S^3$, the eigenvalues should be interpreted as extended objects along the $S^3$, but local in the other directions
of $AdS_5\times S^5$. This is part of the usual behavior of D-branes on compact spaces \cite{WT}.
Indeed, they should be D3-branes, and a single BPS eigenvalue 
becomes a giant graviton wrapping an $S^3$ of $AdS_5$ written in global spherical coordinates. Also, in our case we have a discrete energy spectrum, unlike the continuum energy spectrum of the BFSS model. This is required by the AdS/CFT correspondence, and it implies that the spacetime geometry should have a very different asymptotic behavior which is not asymptotically flat. We still would want to talk of these configurations as a moduli of BPS 
states. This is done if we declare that the hamiltonian is $H=\Delta-3/2 R$ for $R$ a particular $R$ charge which is diagonal in $SO(6)$. This is some twisting of the dynamics by a unitary transformation $U = \exp(i 3/2 R t)$. Since the $R$ charge and $\Delta$ commute, diagonalizing $H$ and $R$ is the same thing as diagonalizing $\Delta$ and $R$. However, with respect to $H$, all the $1/8$ BPS states have zero energy, and the set of classical configurations that satisfy $H=0$ is a manifold, which we will call the moduli space of BPS states.

In our procedure we want to integrate out certain degrees of freedom associated to off-diagonal modes of matrices.
To test wether off-diagonal modes are heavy or not depends on details of the dynamics. For small values of the SYM coupling constant, we need 
very large vevs for the diagonal matrices, large compared to  $1/g_{YM}$ (this produces masses of order one). Factors of $N$ also show up from counting species of off-diagonal elements, so we should use vevs which are generically larger.  The 't Hooft coupling constant is usually 
the leading perturbation parameter, and we need to suppress it by factors of $N$. Therefore the vevs should be comparable to 
$N/\sqrt{\lambda}$ to account for these extra factors of $N$.    Here $\lambda$ is the 't Hooft coupling constant
$\lambda\sim g_{YM}^2N$.  

For intermediate  $\lambda$ coupling, the off-diagonal masses increase with 
$g_{YM}$, so the approximation of commuting matrices to describe the dynamics seems to improve as we tune towards 
large 't Hooft coupling. We are  then allowed to reduce the vevs of the matrices substantially.
For $g_{YM}$ fixed and small (but not tiny), the vevs need to be roughly of order 
\begin{equation}
|\vec \phi|  \sim \sqrt N
\end{equation} 
times a number of order one. We need to check for self consistency of 
of this approximation later on.

   Moreover,  we have a lot of global symmetries between the different matrices. In our problem
   the off-diagonal masses are characterized by the norm of  vectors $|\vec\phi_i-\vec\phi_j|$. We can use the global rotations of the vectors to rotate any of these vector differences to lie along the $Z$ direction. Thus for any relative pair of eigenvalues, this is the same as being half BPS, and in that case we already believe that the dynamics of the eigenvalues is not affected by integrating out other off-diagonal fields. See the appendix \ref{sec:appa} for details on the calculation that shows that to one loop order one gets a zero energy contribution over the classical piece. 
   This can convince us that to one loop order the statements made above are correct.

From this perspective we also find that the volume form associated to the variation $\theta_{ij}$ will have to be equal to 
the length of the three-vector $\vec\phi_i-\vec\phi_j$, which is the same as that for the $\theta_{ji}$ angle as well.
The measure factor associated to the gauge orbit can be nothing other than
\begin{equation}
\mu^2 = \prod_{i<j} |\vec\phi_i-\vec\phi_j|^2
\end{equation}
which looks like a generalized square of the Vandermonde determinant. Except that now the eigenvalues have three components each, and we use the square of the vector differences.
This is also required by the expectation that the result should respect the global invariance under rotations of the configurations.

Notice however, that unlike the case of half BPS states, taking the square root of the measure factor does not lead to different signs for the wave function when we exchange eigenvalues. In the half BPS case, our vectors have only one component, and therefore they are just numbers. The factor $(x_i-x_j)^2$ has an analytic square root given by $(x_i-x_j)$. In the case of higher dimensional vectors,  we are forced to choose the positive branch cut of the square root everywhere if we want a continuous function, so one can not associate the square root of the measure to a fermionization procedure for the eigenvalues. Here the picture starts looking quite different from the one matrix model for half-BPS states, where the fermionization was  natural because it produced smooth wave functions for the fermions without branch cuts \footnote{One can imagine a more complicated form of taking the square root, by noticing that $|\vec\phi_i-\vec\phi_j|^2=\frac 12\tr((\vec \phi_i-\vec\phi_j)\cdot \sigma)^2$, where the $\sigma$ are Pauli matrices. This involves introducing extra degrees of freedom to take into account this spin associated to the $2\times 2$ matrices. It is tempting to speculate that the matrix model fermions will do exactly that, but this type of idea is also reminiscent of twistor constructions. For the purposes of this paper, we will use the naive square root.
}

Given the measure factor, it is now straightforward to write a Hamiltonian for the effective diagonal dynamics. The Hamiltonian is
\begin{equation}
H = \sum _i -\frac 12 \frac 1{ \mu^2} \nabla_i \mu^2 \nabla_i + \frac 12 (\vec\phi_i)^2
\end{equation}
where $\nabla_i$ is the gradient operation associated to the eigenvalue $\vec\phi_i$.
This is the harmonic oscillator Hamiltonian, with a correction factor due to the 
non-trivial change of measure from generic matrices to diagonal ones.

If we take a wave function $\Psi$ in this hamiltonian, and transform it to $\Psi= \psi/\mu$, then probability densities calculated with $\Psi$ in the form $\int \mu^2 \prod d^3\phi_i \Psi^*\Psi$, get transformed 
into probability densities based on the standard measure $\prod\int d^3\phi_i \psi^*\psi$, where we get $N$ copies of the measure associated to each eigenvalue tuple, which are independent of each other.
This wave function $\psi$ can be treated statistically in the same way as our fermionic wave function
for the half-BPS states, because the measure of the eigenvalues can be treated in a regular manner. This change of basis results in a modified hamiltonian.

The modified Hamiltonian for interacting bosons $\tilde H$ is given by
\begin{eqnarray}
\tilde H &=& \sum _i -\frac 12 \frac 1{ \mu} \nabla_i \mu^2 \nabla_i\frac 1\mu + \frac 12 (\vec\phi_i)^2\\
&=& \sum_i \frac 12 \frac 1{ \mu} \nabla_i (\nabla_i \mu)-\frac 12 \frac 1{ \mu} \nabla_i \mu \nabla_i+\frac 12 (\vec\phi_i)^2\\
&=& \sum_i \frac 12 \frac 1{\mu}( \nabla_i \nabla_i \mu) -\frac 12 \nabla_i\nabla_i + \frac 12 (\vec\phi_i)^2\\
&=& \sum_i H_{osc, i} +V_{eff}\end{eqnarray}
 We interpret this as a Hamiltonian for $ N$ non-relativistic particles. Since $N$ is large, we have a lot of identical particles in the problem. This is a regularity that we can handle statistically in a thermodynamic/hydrodynamic  sense .
This change of basis in the wave functions is necessary to be able to give an analog of a droplet picture for these states which preserve less supersymmetry.

In the Hamiltonian $\tilde H$, we have a standard kinetic term and the quadratic potential for each eigenvalue. We also get an effective potential generated by the measure factor. This can be calculated explicitly from the Laplacian of $\mu$. The final answer for the effective potential is that there is  a two body repulsion between eigenvalues, and a three body potential. These particles are to be treated as {\em interacting} bosons. Whether they interact weakly or strongly depends on their separation. If we take them to be very far apart from each other, the 
interactions decay, and the bosons can be treated as free particles. In the ground state, the bosons pile up as close as they can to each other, but then the repulsive interactions dominate. In the ground 
state of the system, the bosons are {\em strongly interacting}: the two body and three body interactions can not be treated as a perturbation. This effect is  due solely to the measure factor: it is an effect of the fact that the theory is gauged. This measure encodes the fact that we have to look only at singlet wave function of the $SU(N)$ dynamics. 

There are two important things we would want to do now. We want to show that this Hamiltonian reproduces the 
energy spectrum of our naive model, and that this formulation captures aspects of the correct
operator algebra of the ${\cal N}=4 $ SYM on $S^3$. This is not obvious anymore, because we can not transform the system to a description in terms of free particles in a simple manner. Indeed, we need to solve what looks like a very complicated quantum system.

\subsection{Description of the quantum states of the matrix model}

Given the hamiltonian $\tilde H$, or $H$, we want to solve the system. What we would like to do is show that we can construct a complete set of wave functions  which are eigenvalues of the Hamiltonian. Barring that complete solution, at least we should be able to get the ground state of the system. 
It turns out that this problem is easier to solve for $H$ than for $\tilde H$. Since we know the relations between the wave functions of both Hamiltonians, we can transform easily between one description and the other one.

The first thing we do, is to notice that the Hamiltonian $H$, apart from the terms associated to the measure factor, looks very similar to a harmonic oscillator. We then try the ground state wave function of the harmonic oscillator as a trial wave function. This, is, we begin with $\Psi_0 \sim \exp(-\sum x_i^2/2)$ and check whether it is a solution of the spectral problem. This turns out to be true.

The simple observation that we need to check this, is that $\vec \nabla_i \Psi_0 = - \vec x_i \Psi_0$. Then
\begin{equation}
H\Psi_0 = -\sum \frac 1{2 \mu^2}\nabla_i ( \mu^2 x_i \Psi_0)+ \frac 12 x_i^2 \Psi_0. \end{equation}
If we let $\nabla_i$ act on $\Psi_0$ first, we see that we cancel the quadratic term in $x_i^2$ from the potential. The question of wether we get an eigenvalue of the energy reduces to calculating if for the remaining equation we get a pure number in the right hand side of the following equation
\begin{equation}
\sum \frac 1{\mu^2} \nabla_i (x_i \mu^2)= \kappa
\end{equation}
We  now use the fact that $\vec \nabla_i \vec x_i = d+\vec x_i\vec \nabla_i$, where $d$ is the number of matrices (we are using the identity on vectors in dimension $d$ using vector notation) to find
that the trial wave function is an eigenvalue of the Hamiltonian if and only if $\mu^2$ is an eigenvalue of
\begin{equation}
(\sum \vec x_i\cdot\vec \nabla_i)
\end{equation}
But we recognize this operator immediately: it is the generator of infinitesimal transformations for  scaling transformations where all $x_i$ are scaled by a uniform common amount. The function $\mu^2$ is an eigenvalue if it is a homogeneous function. Indeed, $\mu^2$ is a homogeneous polynomial on the eigenvalues of degree $N(N-1)$, which we have calculated explicitly. The eigenvalue problem we started  with is in the end reduced to counting the degree of the polynomial measure.
We find that for different dimensions associated to the bosons,
 we have found at least one eigenvalue of the 
energy. This should be identified with the ground state wave function.

Now, let us look for excited wave functions. The analogy to the harmonic oscillator combined with the appearance of the scaling operator suggests that they should all be given by polynomials of the $x_i$ variables, $P$, times $\Psi_0$. If this is true, then we can reduce the problem of the full set of wave functions to properties 
of polynomials which can be checked degree by degree, starting from the highest to the lowest. We can now check this type of ansatz explicitly.

A proof would proceed by starting with the term with the highest (multi)-degree in $P$. It is easy to show that the mathematical manipulations closely follow the ground state if we let $\nabla_i$ act on $\Psi_0$ first multiple times. This cancels the quadratic potential term.
To subleading order, if we let only one of the $\nabla_i$ act on $\Psi_0$, we again get the homogeneity operator, and if the wave function is an eigenvector of the Hamiltonian, the energy has to measure the degree of the polynomial. The fact that polynomials would not just be homogeneous polynomials follows from the fact that
we can get additional terms where no derivative acts on $\Psi_0$. The spectral problem 
gives us a relation between the subleading terms in $P$, double derivatives of $P$ and derivatives of $P$ multiplied by $ \mu^{-2} \partial \mu^2$. The double derivative term automatically gives a 
polynomial, so we don't have to worry about it at this moment. However, we also have the terms with a derivative of $P$ times a derivative of $\mu^2$, which is multiplied by $\frac 1{\mu^2}$.  These terms are usually problematic.

 We can now do an explicit evaluation of $\vec A_i=\mu^{-2} \nabla_i \mu^2$.
This is equal to
\begin{equation}
\vec A_i = 2 \sum_{j\neq i} \frac{(\vec x_i-\vec x_j)}{ |x_i-x_j|^2}= 2\sum_{j\neq i} \vec A_{ij}
\end{equation}
Notice  that there are potential cancellations due to the fact that the wave function is symmetric in the exchange of the $x_i$, and the $A_{ij}$ is antisymmetric.

The schematic equation we would need to solve for $P$ is of the form
\begin{equation}
\sum A_{ij}\dot \nabla_i P +\nabla_i^2 P \sim P_{sub} 
\end{equation}
The terms with $A_{ij}$ and $\nabla$ reduce the degree of the terms in $P$ by two.
In the notation above $P_{sub}$ is the subleading piece of the polynomial $P$. 
For the case of $d=1$, the fact that $P$ is a symmetric wave function is enough to show that
$A_{ij} (\partial_i P -\partial_j P)$ is a polynomial, because $(x_i-x_j)$ is a factor of $(\partial_i P -\partial_j P)$, as this function is antisymmetric in the exchange of $x_i\leftrightarrow x_j$ and vanishes exactly when $x_i=x_j$. 

It is easy to show by direct calculation that for linear symmetric 
polynomials there is no problem for any $d$.
 Indeed, we can define the d variables
$t_\alpha =\sum x_i^\alpha \sim \tr( x^\alpha)$. If we consider a polynomial of the $t$ variables alone, then 
$\partial_{x^\alpha_i} \sim \partial_{t_\alpha}$ on this restricted set.  We see that then $\vec A_{ij}$ ends up multiplied by $\nabla_t-\nabla_t\equiv 0$. Thus we can generate all polynomials in the $t$ this way. This is analogous to the center of mass motion of the 
system. This center of mass motion decouples. This is why it is easy to find these wave functions.

This set of polynomials is very restricted. They should be identified  with quanta of the trace part of the matrices in the ${\cal N}=4 $ SYM theory. We see that we are making progress in the right 
direction. At degree two, we start getting trouble to solve the equations for $d>1$, but one can find at least one rotationally invariant solution whose leading polynomial term is $\Omega= \sum \vec x_i^2$.
Again, one can find that all polynomials of $\Omega$ form a subset of the eigenspaces.

Even though the polynomials don't seem to give a complete solution to the dynamics, this should not bother us. We don't believe the dynamics above is exact anyhow. Moving to phase space, instead of working on the position space does not change substantially the type of analysis we are doing. We should simply replace the measure factor by a similar term which involves $x$ and $p$ variables at the same time.

The advantage of working in phase space is that we should be able to describe the wave function by holomorphic polynomials of the variables $X,Z$ or $X,Y,Z$. This is a choice of complex polarization for our wave functions. In the one dimensional case, the wave functions are homogeneous polynomials. They are normalized with a measure factor given by $|\Delta(Z)|^2 \exp(-\tr(Z\bar Z))$. Here $\Delta(Z)$ is the Van-der-Monde determinant for
the matrix $Z$.

For one eighth BPS states, we should get the same type of result. We have already seen that
polynomials in $x$ space multiplied by the Gaussian factor sometimes give 
wave functions which are eigenvalues of energy. We will {\bf conjecture} that for one quarter and one eighth BPS states, all the wave functions that diagonalize the true Hamiltonian are described by homogeneous polynomials of $2N$ or $3N$ complex variables respectively,  organized into $N$ two vectors or three vectors  of complex variables (a dimension 2 or 3 complex vector space), which are symmetric under the exchange of any pair of vectors. These are going to be accompanied by some measure factor
for normalization.
The associated measure should be given (at least to leading order) by
\begin{equation}
\int \prod_i d^d\phi_i d^d\bar\phi_i \prod_{i<j}| \vec \phi_i-\vec\phi_j|^2 \exp( -\tr(|\phi|^2))\label{eq:measure}
\end{equation}
If we absorb the square root of the measure factor into the wave functions, we have $N$ copies of the canonical measure for the $\vec \phi_1$. And in this case we interpret the square of the wave function probabilistically along the same lines as in in the section on half-BPS states.  Indeed, the type of answer we are giving here is the exact analog as that for half-BPS states. The wave functions are described by some free bosons, convoluted with some ground state wave function which takes care of all of the measure factors.

Because the wave functions are related to those of free bosons, the counting of states and degeneracies of BPS states (as well as their representations under the $SU(2)$ or $SU(3)$ symmetries that relate the quarter BPS states amongst themselves) will agree with what we have found already in the naive model. 

Notice that up to this point, we have not really solved the effective dynamics from first principles. Instead, we have made various arguments that make the final answer we wrote very plausible. Also notice that in some sense our approach works best at strong 't Hooft coupling.
We have balanced it with moduli directions so that we can treat the problem perturbatively.
This is analogous to going semiclassical by requiring very large quantum numbers. 
The type of argument we are doing is to be considered as a {\bf strong-coupling expansion} of ${\cal N}=4$ SYM around certain supersymmetric configurations.

The correct answer should not be too far from what we have tried to argue. Maybe there are additional terms in the measure that need to be included when we take into account
corrections in $\lambda^{-1}$ etc. These are beyond the scope of the present paper but they should be analyzed systematically.

This description of the system is still given in terms of very few variables of the original dynamical system. These are the moduli variables, and we have a description of all the states we want to consider in terms of the wave functions of the moduli fields.

\section{Hydrodynamics: the brane description of  BPS states.}\label{sec:saddle}

We arrived in the previous section at a place where we are allowed to treat our problem of quarter or one eighth BPS states statistically. We have identified a candidate dynamics where we can write all the microscopic wave functions in some preferred basis, and the variables used to describe the
dynamics have a lot of regularities that can be treated statistically.

The dynamics is characterized by $N$ particle wave functions, where each of the particles live in a two  or three dimensional complex flat space (a four or six dimensional real vector 
space associated to two or three  coordinates and  momenta per particle). Label these collectively 
by a  vector $\vec x_i$. These particles are interacting bosons which repel each other.

The ground state wave function (with energy zero) in terms of the $\vec x_i$ is given by
\begin{equation}
\psi_0
\sim \sqrt{\prod_{i<j} |\vec x_i-\vec x_j|^2}\exp(-\sum_i \frac {\vec x_i^2}2)
\end{equation}
If for each $\vec x$ we define the two complex variables $z^1 = x^1+i x^2$, and $z^2= x^3+ix^4$, $z^3= x^5+ix^6$ which can be organized into a complex  vector $\vec z_i$ for each eigenvalue.

All other wave functions, of energy $n$, are produced by multiplying the ground state wave function by a homogeneous polynomial of degree $n$ of the $\vec z_i$ which are invariant under 
permutations of the particle labels. This is the most important dynamical statement we can make: we have a candidate description that solves the dynamics in terms of 
wave functions we can calculate.

The square of the ground state wave function is given by
\begin{eqnarray}
|\psi_0|^2 &\sim & \prod_{i<j} |\vec x_i-\vec x_j|^2
\exp(-\sum_i  {\vec x_i^2}\\
&\sim&\exp\left(-\sum_i {\vec x_i}^2 + 2 \sum_{i<j}\log (|\vec x_i -\vec x_j|)\right)
\end{eqnarray}

Again, as in section \ref{sec:halfbps} this formula has a probabilistic interpretation as some type of Boltzman 
distribution for a gas of $N$ particles with logarithmic repulsive interactions (long range repulsion) in four or six dimensions which are confined by a harmonic oscillator well. The system is at some fixed temperature. 

The statistically dominant configuration of the system is given by some configuration which can be described by a density of particles on the plane $\rho(\vec x_i)$, which should be close to the lowest energy configuration of the system. The energy function in terms of this density is given by
\begin{equation}
E \sim \int d^4 x \rho(x) x^2 -  \int d^4 x d^4 y \rho(\vec x)\rho(\vec y)\log(|\vec x -\vec y|)
\end{equation}

The function $\rho$ is constrained by $\int d^{4,6} x \rho(x) = N$ and by $\rho(x) \geq 0$. 
Describing the system in terms of a density of eigenvalues is a coarse grained approximation to the problem.

We can find the minimum energy configuration by a variational method. The constraint $\rho(x)\geq 0$ is taken care of by saying that the variation of the energy functional with respect to $\rho$ is different from zero only on the domain of integration.

A straightforward variation shows that on the domain of integration we should have
\begin{equation}
x^2 +C - 2 \int d^4 y \rho(\vec y) \log(|\vec x -\vec y|)=0 \label{eq:var1}
\end{equation}
This is an integral equation that determines $\rho(x)$. Again, $C$ is a Lagrange multiplier enforcing the constraint on the total number of particles.
 If $\rho (x) \geq 0$ 
is non-singular at $x$, then we can take derivatives of equation \ref{eq:var1}. Indeed, 
\begin{equation}
\nabla_x^2 \log(|\vec x -\vec y|) \sim \frac 1{|\vec x-\vec y|^2}
\end{equation}
and $\frac 1{|\vec x-\vec y|^2}$ is proportional to the Green's function of the Laplace operator in four dimensions. From here, it follows that for quarter BPS operators
 \begin{equation}(\nabla^2 )^2 \log(|\vec x-\vec y|) \sim \delta^4 (x-y)\end{equation}
With this we find that $\rho(x)=0$ for such points.
In the one eighth BPS case, we use the fact that $\nabla^3 \log|x-y| \sim \delta(x-y)$
to reach the same conclusion.

From this calculation we find that the distribution of particles on the four or six plane is singular. Because of spherical symmetry, we can assume that all particles are located uniformly on a very thin shell around $r_0$ so that $\rho(\vec x) \sim N  \delta(|\vec x|-r_0)/r_0^3$. This is the simplest singular behavior we can imagine. It would be very useful to show that this is the 
case in general. 
By doing a force argument, the repulsion of the particles should balance the confining 
force.  Since the typical distance between the particles is characterized by $r_0$, the 
repulsion should scale like $N r^{-1}_0$, while the confining force is of order $r_0$. This gives us a scaling whereby $r_0 \sim \sqrt N$. This matches qualitatively the condition that we
had imposed on the energies of off-diagonal matrix model  components for our approximations to be valid. In this sense, this provides a self-consistent description of the dynamics. 

 We should think of this distribution of particles 
in the quarter BPS case as some type of three-dimensional  brane in the shape of a round $S^3$, or as some type of fluid confined to an $S^3$. In the case of one eighth BPS states, the shape of this eigenvalue brane is a round $S^5$. These geometries associated to  the eigenvalue distributions will be called 
E-branes, to distinguish them from D-branes in the geometry. 

Notice that the typical separation between any chosen random two particles grows like $\sqrt N$ when we let $N$ be large. Remember that in the quantum mechanical model we have a preferred length scale determined by $\hbar$. This means that this E-brane becomes macroscopically large
with respect to $\hbar$, and the system should begin to display classical features. Indeed, the growth with $N$ of the shape scale factor is exactly the same as for the case of the half-BPS droplet.  Thus the size of the E-brane should be interpreted as the radius of of the $AdS_5\times S^5$ geometry. Here we are already seeing the appearance of a round 
$S^3$ for quarter BPS states, and we get the full $S^5$ for one eighth BPS states, so we are starting to get closer to the full spacetime geometry. If we understand the geometry of the AdS/CFT correctly, there are four directions of $AdS_5$ that are already part of the field theory description.
These are the angles of the $S^3$ boundary at infinity, and the time direction. We are only 
missing the radial direction. Moreover, we have an $S^3$ or $S^5$ generated from the dynamics which we want to identify with the corresponding geometric slice of the physical $S^5$ in the AdS geometry.

One needs to be careful: the statistical model is $SO(4)$ or $SO(6)$ symmetric, however the quantum mechanics we have described is not. Indeed the complex structure can be associated to a symplectic form on phase space. The pullback of this form to the worldvolume of this E-brane is not
zero. This means the particles can be treated as  being charged in some type of magnetic field, and transport of particles can only happen along magnetic field lines. The motion transverse to these lines is confined due to magnetic effects. 
These magnetic field lines foliate the three-sphere along the Hopf fibration. This is very similar to the quantum hall effect in higher dimensions \cite{PE}.  This can be 
drawn as in figure \ref{fig:membrane.eps}

\myfig{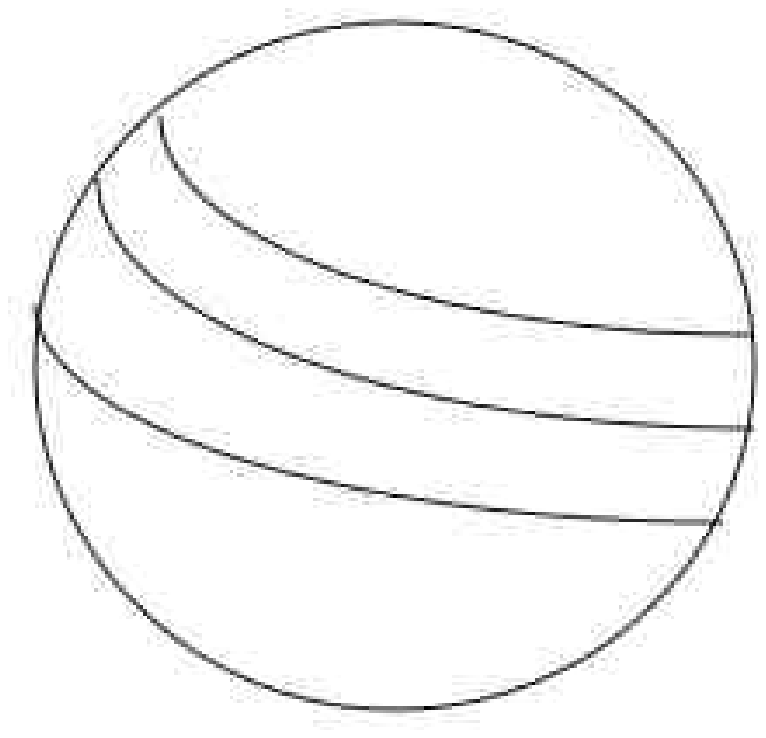}{5}{The quarter BPS membrane with the Hopf fibration: the curves are
BPS trajectories on the E-brane,  and motion happens along them at the speed of sound.}

In many senses this result is natural. After all, we are studying only quarter BPS states with 
respect to some supersymmetry. The quarter BPS gravitons for some associated R-charge 
will all flow along a geodesic on a particular $S^3$ equator of $S^5$, and they will do so
along the direction specified by the vector field on the sphere which rotates the configuration by $R$ transformations. This is done along a Hopf fibration of the $S^3$ which is adapted to the $R$ charge associated to $1/4$ BPS states. There is a particular $J$ associated to an $SO(2)$ rotation in $SO(6)$ which is a symmetry of all these configurations. This $SO(2)$ 
has fixed points on the $S^3$ manifold that we associated above for quarter BPS states.
This fibration structure suggests a close connection to the way geometries are described in the half BPS case \cite{LLM}, where the inside of the $S^3$ E-brane can be associated to a degeneration of the $SO(4)$ isometries of the boundary, while the outside of the E-brane is associated 
to the locus of this $SO(2)$ degeneration. This would be an analogous 
construction of boundary data to the LLM geometries.

We can do the same analysis for the $1/8 $ BPS states. The equivalent of the Hopf fibration is the statement that $S^5$ is a circle bundle over $\CP^2$. In this case, the circle never shrinks to zero size on the $S^5$, and the circles are all diameters of the $S^5$. This fibration is  associated to a vector field on the $S^5$ along which the excitations flow. Because there are no obvious degenerations, we expect that an LLM type of reconstruction of the geometry will work differently.

Now, to understand the dynamics of this system and to be able to compare it with supergravity,
 we need to look at the spectrum of excitations of the model
and the corresponding geometrical picture.

This is easy to do for BPS excitations. Again traces of polynomials in $z^1, z^2, z^3$ should provide the variables
that correspond to gravitational quanta \cite{AdSCFT2}. Now we follow the same type of analysis done for half BPS operators: we multiply our wave function by a truncated coherent state with small 
parameter (once we normalize the traces correctly), this is, we consider
\begin{equation}
\ket{\alpha_{nm}}\exp( \alpha_{nm} N_{n,m}\tr((z^1)n(z^2)^m))\psi_0
\end{equation}

The trick is again the same as done above: 
write $|\psi_0|^2$. Take the exponential term in the coherent state wave function and  treat
it like a small correction to the background potential. This will change the balance of forces, and the shape of the E-brane will change to compensate it. The  potential one gets this way
 is not the most general potential, because the wave function is holomorphic. This also shows that these deformations don't diffuse the E-brane: the E-brane should still be infinitesimally thin. The argument is along the same lines as when we proved that the eigenvalue distribution is singular for the ground state configuration. This is because the Laplacian (squared or cubed) acting on a holomorphic function is zero, so the particle distribution density can not have smooth support. This is a very important point when one wants to embed the E-brane geometry in the AdS spacetime. \footnote{This is similar to stating that in the half BPS geometry the classical configurations are incompressible, and the edge of the droplet is sharp.}

  The potential for the individual eigenvalues is then given by the real part of the exponential term in the coherent state.
If we let time evolve, then different terms in the coherent state expression will have different energies which are shifted by integers. This makes the shape of the perturbation evolve uniformly in time, as these changes can be absorbed into a phase rotation of all the $z_i$.  
In essence, the initial shape of the geometry is kept fixed up to some trivial rotations of the configuration.

This mimics precisely the description of half BPS states. Again, we have a constant speed of sound associated to these shape perturbations,  which we want to identify with the speed of light
on the supergravity geometry. Moreover, we have as many deformations as there are BPS gravitons with the given quantum  numbers. This follows from the AdS/CFT correspondence established in \cite{AdSCFT2}.

We see that our calculations have provided for us a geometric object with some peculiar hydrodynamics on it. The hydrodynamics is such that in this sector the collective motion of particles on the membrane associated to sound, coincides with motion of gravitons at the origin of $AdS_5$ which lie on
an $S^3$ equator and which are BPS with respect to a particular R-charge, or which follow a particular fibration of the $S^5$ geometry with respect to some R-charge.

It is natural then to conjecture that the $S^3$  we found in phase space embeds itself in spacetime in some particular way. This is to be interpreted in the same way than the edge 
of the half BPS droplet embeds into a particular diameter on the $S^5$. Given our geometric object in phase space, there seems to be an internal and an external region to 
it, which should also have some interpretation in spacetime. One can hope that these denote some degeneration locus of some particular fibration, in a similar spirit to LLM.
Indeed, there is an extra $U(1)$ R charge in the Cartan of $SU(3)$. This $R_3$ degenerates on the locus we described. Since this is associated to a conserved $U(1)$ isometry of the
geometry, the surface we are describing can be associated to this particular 
degeneration.  Ten dimensional spacetime should then be reconstructed by solving some partial differential equation, whose boundary data is the 
membrane configuration in $\BC^2$ that we have 
found. We will call this spacetime the LLM transform of the hydrodynamic configuration for quarter BPS states.

The important point of this transform is that the locus where the 
eigenvalues are located corresponds to a particular cross section of the geometry, and for BPS motion restricted to that locus, there 
is a speed of light that we can associate to the geometry and which matches the calculations we can do in our coarse grained setup.

Now, regarding the $1/8$ BPS problem, one can imagine that something similar
to what we have done might work. We have found  an $S^5$ in a six dimensional phase space of radius of order $\sqrt N$. This would be a derivation of the $S^5$ part of the geometry of $AdS_5\times S^5$ from first principles. We want to identify this $S^5$ with the one given in the geometry.

However, in this case there seems to be no room for an LLM transform of the geometry as argued above for the quarter and half BPS states. 
In this case it is  not clear what the inside of the sphere represents in spacetime. This is because there is no additional unbroken symmetry that one could say degenerates at that 
locus: the $SO(4)$ isometry is only supposed to degenerate on an $S^5$.
Instead, we should take a different approach to understand the full geometry in that case.
This extra direction that one needs to deal with will be the $AdS$ radial direction.
To understand it well one probably also needs to understand the supergravity modes that propagate there better. This involves including the higher spherical harmonics of the fields in the SYM. This is beyond the scope of the present paper. Instead, we will look at other 
features related to writing different configurations of branes and topology changes.

\section{D-branes and topology changes in  BPS geometries}\label{sec:top}

We have already described the ground state of $AdS_5\times S^5$ and 
the coherent perturbations about it. It is natural to presume that if we can go further and describe topology changes in the half BPS case, then we can do the same for the $1/4$ and $1/8$ BPS case as well. Indeed, the way topology changes were understood in \cite{LLM} was as geometric transitions when various branes were stacked on top of each other.

Here, we want to proceed along the same lines. In the half BPS picture, the two natural D-branes
associated to giant gravitons were described either by particles or holes in the free fermion picture \cite{toyads}. The eigenvalues far away from the droplet were interpreted as giant gravitons growing into $AdS_5$. The holes were interpreted as the giant gravitons that grow in the $S^5$.
 In the quarter BPS and one eighth case we don't have free fermions, but we do have particles, so at least the eigenvalues far from the brane can be constructed.
 This should correspond to a D-brane just like in the half BPS case. If we stack a lot of these eigenvalues on top of each other, and separate the branes very much, then we can have topologies with various collective geometries on the quantum 6-plane. Thus, what we called the 
 E-brane might consist of various disconnected pieces. The number of such connected components will presumably be measuring some Betti number of the geometry.

It is harder to understand what is one supposed to do about states that correspond to giant gravitons growing into the $S^5$. There are various reasons why this is harder to understand. These objects should become extended on the $S^5$, and can not be considered as point particles on phase space. Moreover, the shape of these extended objects can vary a lot, and in general one can relate them to holomorphic surfaces on the a cone over the $S^5$ \cite{Mikh}. Indeed, this forces us to consider a
problem with a lot of moduli.

We will use the half BPS case as inspiration to describe these objects. Indeed, the half BPS
hole wave functions should be reproduced in some form from our arguments. The only difference between a hole wave function and the ground state was that we multiplied the ground state wave function by a determinant operator, of the form 
$\psi\sim\det(Z-\lambda)\vac$, and for $m$ holes  on top of each other, we use $\det(Z-\lambda)^m$. We will do the same here, at least for the half BPS case and let us see what happens.

Indeed, we need to go to our saddle point description, and calculate the saddle point
of $|\psi|^2$ with the new wave function.
The determinant has a simple logarithm of the single trace  form 
\begin{equation}
\log(\det(Z-\lambda))=
\tr\log(Z_\lambda)\sim \int \rho(z_1,z_2, z_3) \log(z_1-\lambda)
\end{equation}
This can be interpreted as a deformation of the confining potential for a single particle. 
Indeed, with the anti-holomorphic piece added from calculating $\log (|\psi|^2)$, 
we get an effective repulsion from the full hyperplane $z_1=\lambda$.

This hyperplane intersects the $S^5$ along a three sphere. This intersection manifold is located exactly where we would expect the giant graviton to be: the intersection of a holomorphic surface with the $S^5$.

If we stack a lot of these branes on top of each other, the hole that they tear on the saddle point E-brane should become macroscopically large. One can wonder then if one should consider 
topologies of where the E-brane ends. This seems unlikely. Since the particles in the E-brane repel each other, and if they are also repelled from the $\lambda$ plane, this would generally give rise to an unstable equilibrium where the end of the E-brane can move elsewhere. Instead, it is natural to propose that the geometry becomes a five dimensional doughnut, as in the figure \ref{fig:doughnut.eps}

\myfig{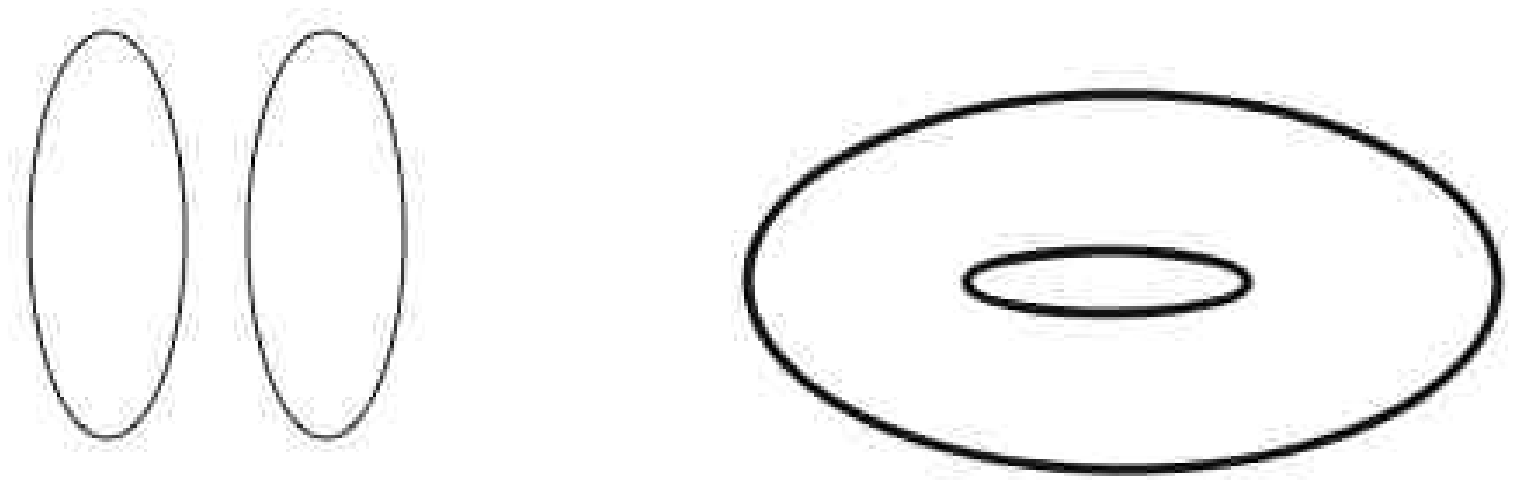}{6}{Topology changes of the eigenvalue distribution due to condensation of many giant gravitons. Two different cross sections are shown}

Also, this construction suggest a way to understand more general giant gravitons. Indeed, the connection to holomorphy provides the key. If we want some giant graviton associated to some hypersurface on $\BC^3$ which is given by a polynomial in three variables $f(z_1,z_2,z_3)$, it is natural to multiply the wave function by
\begin{equation}
\psi_f \sim \det (f(X,Y,Z)) \psi_0
\end{equation}
Again, in the saddle point description, we have to replace sums over eigenvalues by integrals,  so that $\log\det(f(Y,Z))$ should be replaced by 
$\int \rho(z_1, z_2,z_3) \log(f(z_1, z_2,z_3))$. 

This again produces a repulsive force from the hypersurface determined by $f=0$, which is where the logarithm becomes singular.  For maximal giant objects, the curves are supposed to be holomorphic 
on the base of the Hopf or $U(1)$ fibration over ${\CP}^2$ and extend uniformly along the  fiber. This is accomplished by $f$ that are homogeneous functions of $z_1, z_2,z_3$, as these are invariant under simultaneous rotations of $z_1, z_2,z_3$.  The degree of $f$ then measures the number of branes. If $f$ has repeated roots (or if it is an $m$-th power of some smaller polynomial), we end up with various branes on top of each other. \footnote{This discussion only covers the  $1/8$ BPS case for states where we don't turn the fermions on. It would be interesting to understand the more general case which involves fermions too. 
It has been suggested in the literature that there are $1/8$ BPS D-branes which have electric and magnetic fields on their worldvolume \cite{Kor}, and these should provide the additional states that are required to match $1/8$ BPS states with fermionic variables turned on.}

Now, we can try to consider what type of topology changes are affected by many giants on top of each other in this more general setup. Indeed, we would expect the E-brane to still be a continuous object, but that avoids the holomorphic  hypersurface where the function $f=0$.
We should  expect the E-brane to form some tube around $f=0$ which stops where $f$ intersects the $S^5$ and opens up into the $S^5$. A cartoon of this  tube is presented in figure
\ref{fig:doughnut2.eps}.  This means that topologies will in general be a lot more complicated in
the case of quarter BPS and one eight BPS geometries than in the case of the half BPS case.

\myfig{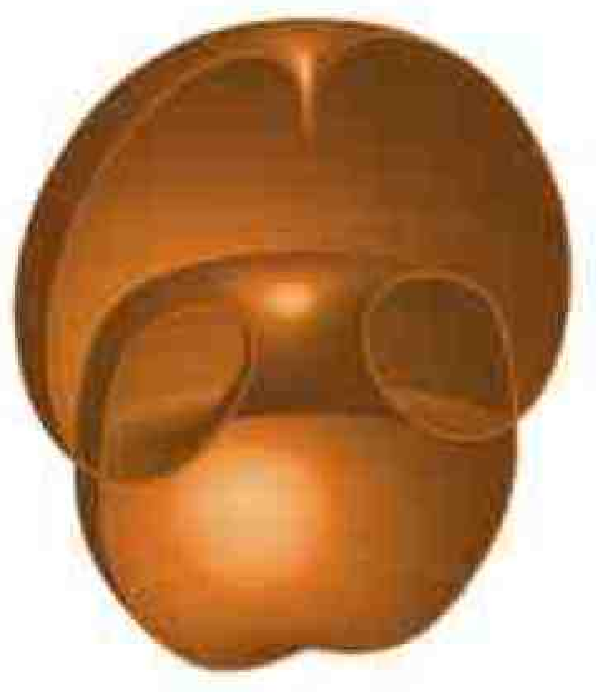}{4}{A tube around the function $f=0$ which ends in a knot on $S^3$. The surface drawn is $f=0$, slightly thickened. The surface does not self-intersect in higher dimensions. This is an artifact of the two dimensional visualization}

One should be able to say something more precise about these topologies. In general one can imagine that the E-brane surfaces divide the phase space into two regions (inside the E-branes and outside them). It also seems to be hard to put one E-brane inside another one. The configurations that we have described seem to either open 
a tube along some holomorphic curve, or produce new E-branes far away from 
each other.
We will conjecture that this is always so. It would also be convenient if the branes would follow equipotential surfaces  of a one particle potential which is the holomorphic plus antiholomorphic
 plus the harmonic oscillator term. This seems to give the same type of
topologies as we have been describing. It would be very interesting if something like this could be proved.

In any case, the half BPS configurations can be obtained by projection of the E-brane 
surface to a complex line when there is the additional $SO(4)$ symmetry in the other directions. The fact that E-branes can not be nested inside each 
other  translates into having a distinct filled droplet shape for each E-brane 
component. This is also one of the reasons why we believe that E-branes can not be nested.
In some sense this casts doubt on the ``inside region" of the E-brane having meaning in 
supergravity for the case of the $1/8$ BPS geometries: it seems that one can only access it with D-branes by deforming the E-brane shape sufficiently.

\section{Towards strings on $S^5$ and the string scale}\label{sec:string}

So far we have studied only BPS objects in ${\cal N}=4 $ SYM theory on $S^3$. We have found that BPS configurations have moduli that are associated to having vevs in the quantum mechanical system for three complex matrices that commute, which is 
similar to the moduli space of vacua of the ${\cal N}=4 $ SYM theory on flat space. This is characterized by the $s$ wave modes of $X, Y, Z$ to be commuting matrices.

If we start with a single trace state which in the free field theory limit is $1/4 $ BPS, then it must be a trace made of the operators $Y,Z$ in some order. Also, traces of small numbers of letters are identified with strings, so if we want to find massive string states, this is a good place to start.

If the state is going to become non-BPS when we turn on the interactions, then the state must be such that it turns on non-trivial commutators for the matrices. Indeed, the potential of the theory is of the form $\tr([Y,Z][\bar Y,\bar Z])$, which gets contributions from non-trivial vevs of commutators. The lowest lying state of the $SU(2)$ sector which is non-BPS is given by the 
following single trace operator 
\begin{equation}
(\tr(YZYZ)-\tr(Y^2Z^2)\vac\sim\frac 12\tr([Y,Z][Y,Z])\vac 
\end{equation}
We notice that this operator is a square of a commutator. The BPS single trace operator with the same
number of $Y, Z$ is given by $2\tr(Z^2Y^2)+\tr(YZYZ)$. This is obtained by $SU(2)$ symmetry generators  acting on $\tr(Z^4)$. It is also easy to show that these are orthogonal.

In order to turn on a non-trivial commutator, first we need to distinguish two eigenvalues with respect to which we are turning on off-diagonal terms. To do this, we first need to give some energy to these eigenvalues to try to separate them from the ground state E-brane. Clearly it is easier to identify commutators when vevs are sufficiently large. However, we don't want to go to the regime where these eigenvalues become non-perturbative objects, where the eigenvalues become D-branes and the off-diagonal terms become strings stretching between them. This requires an energy which is less than of order $\sqrt N$. We also have to worry about identifying commutators in the strong coupling limit.
This can be done if the energy on the diagonal pieces is sufficiently large, as then the collective effects of the BPS E-brane can be systematically ignored, and the effective $N$ for the calculation just involves the number of eigenvalues with respect to which the commutators are non-zero.
The optimal place to be is at the transition point between the physics being dominated by D-branes and strings. This is the place where we will begin our investigation of string states.

Turning on energy for eigenvalues can be done in the BPS regime, and for sufficiently high energy as we are requiring, of order $\sqrt N$, we end up focusing on the plane wave limit \cite{BMN}. 

We want to have now a picture of how strings become local on the membrane geometry, and that there is a new scale, the string scale, which dominates the dynamics of commutators.

Let us say that we take a state with a lot of energy on $Z$, and a couple of off-diagonal quanta in $Y$. If we have $k$ off diagonal quanta, we should involve at most $k$ different eigenvalues of $Z$. This means that the energy on $Z$ should mostly be associated to these $k$ eigenvalues, and let us say that they are $z_1, z_2, \dots z_k$. Turning on the commutator
terms produces energies of order $g_{YM}|z_i-z_j|$ from the semiclassical description of the Lagrangian. However, in the ground state the eigenvalues are not at zero. This is the consequence of the droplet picture we have been arguing about. The eigenvalues are repelled from each other by a quantum mechanical measure term. Indeed, the $z_i$
are of order $\sqrt N$ in size, and they can not stray very far from the place they begin in the 
$S^3$ as we excite them because the forces become large on the eigenvalue by doing that.

We can picture a non-trivial commutator between different eigenvalues by drawing a straight arrow between them in the droplet picture.
 Then, Gauss law for the $U(1)^k$ commutant group of the $z_i$ requires that the number of arrows incoming and outgoing from each eigenvalue are identical. We can thus form a loop
 by following the arrows. Indeed, if we connect $k$ distinct eigenvalues, there is going to be a unique path to follow. These eigenvalues do not have enough energy to be separated from the droplet, but they have been distinguished by having extra energy associated to them, and some angular momentum quantum numbers determined by the $z_i$.
 This is depicted in figure \ref{fig:clstring.eps}. Each one of the segments will be called a {\em string bit}. This is a  different notion than the way we usually associate string bits to each letter symbol in a word when we describe the spectrum of operators in SYM theory.

\myfig{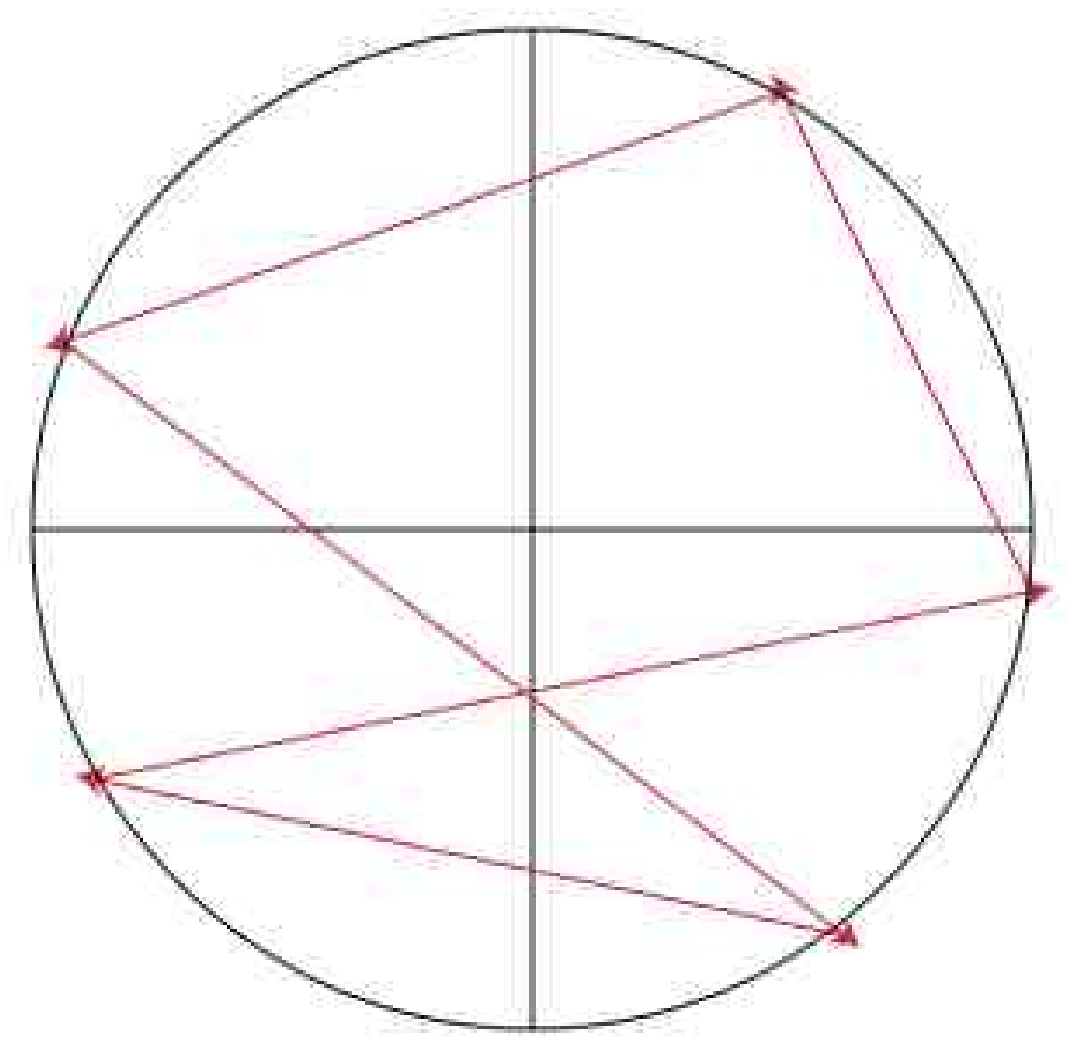}{5}{A picture of a closed string in the droplet picture. Eigenvalues are
distinguished on the circle where they have acquired their quantum numbers, and they are joined by string bits}

Now, we will start worrying about energetics of the string state. Indeed, the figure \ref{fig:clstring.eps} is not at all matching our intuition for what strings do in a classical geometry.
This is because we have not discussed the energetics associated to string configurations. 
At weak coupling the interaction term that produces masses for 
off-diagonal pieces becomes almost irrelevant, and configurations with long string bits have almost the same energy as configurations with small string bits.
Indeed, the commutators give a mass term for off-diagonal terms of order $g_{YM}|z-i-z_j|$ for each segment. If we have finite energy, and we rescale units to the size of the three sphere, we get that $z_i \sim r_i \sqrt N$ and therefore the energies of the string bits are of order
\begin{equation}
E_{ij} \sim \sqrt{1+\lambda |r_i-r_j|^2}
\end{equation}
so at weak 't Hooft coupling, long and short string bits carry roughly the same energy.

However, if we let $\lambda$ be large, which is the regime for the $AdS/CFT$ correspondence to be described by a classical gravitational background, then the off-diagonal energies are of order
\begin{equation}
E_{ij}\sim \lambda^{1/2}|r_i-r_j|
\end{equation}
 If we have finite energy to make a string, and $\lambda$ is large, it must follow that $|r_i-r_j|$ has to be very small. For two $r_i$ on the surface of the sphere of radius $\sqrt N$, this implies that they are very close to each other, and the segment joining them becomes essentially tangent to the sphere. Notice moreover that since the string bits cost energy proportional to the distance on the $S^5$ to a first approximation, we start seeing 
effects associated to the string tension and the string scale.

Notice that longer string bits and various shorter string bits joining two eigenvalues give energies
which are not too different from each other. This is because energy is measured by distance.
 At strong coupling we expect all these states to mix with each other, so that the total number of string bits is not a good quantum number.
Also, there is a lot more entropy associated to short string bits than to longer string bits for a given energy. This mixing of states then implies that the dominant configurations that we identify as strings are made of a lot of small string bits.
In this way we can start thinking of semiclassical strings being given as curves on $S^5$,  as objects which join a lot of eigenvalues so that the string state can be well approximated by a continuous curve.

So far our discussion has centered on identifying some special eigenvalues and constructing the strings as if they were given by adding bits that stretch between different D-branes. In some sense, these eigenvalues are dissolved in the E-brane, so the energy that we associate to
them could also be considered as BPS energy resulting from collective effects on the ground state E-brane we have been describing. This is, the energy associated to the eigenvalues can be replaced by hydrodynamic excitations of the E-brane. In our string bit picture, this is to say that the string
bits gets hydrodynamically dressed. As seen from the point of view of the string  in $AdS_5\times S^5$, this is to say that the string bits are gravitationally dressed, and we can not talk about a string without realizing that it deforms the geometry too. This suggests that strings interact with gravity just like gravity itself, and  that the universality of gravitational interactions is a property of the strong dynamical gravitational dressing
of all states. This could become a proof of the equivalence principle for gravity from the CFT dual.

Being more precise about the string scale is obviously difficult and it is beyond the scope of the present paper, at least at the level of short strings.  Nevertheless, it is an important test of the AdS/CFT duality. It is possible however to calculate the string tension. This is because long strings will have an energy which scales as
\begin{equation}
E \sim T_{string} {\ell }\sim \ell/l_s^2
\end{equation}
where $\ell$ is the length of the string. Since these energies are proportional to
$\lambda^{1/2} \ell $ where $\ell$ is measured in $AdS$ units as shown above, one finds the usual result \cite{M} that 
\begin{equation}
l_s \sim R /\lambda^{1/4}
\end{equation}

 Also, our arguments in this section have been very heuristic. It would be worthwhile to make the statements written above precise. In particular one would want to understand how the repulsion between eigenvalues, and the cost in energy for producing string bits compete with each other. 

Also, one would want to understand the splitting and joining of strings. In some sense the description above in terms of dressed string bits lets the string bits jump between different eigenvalues maintaining the constraints of the system. The splitting and joining would proceed by having four string bits meeting at an eigenvalue and changing the way we follow the 
paths. This would imply that strings interact locally on spacetime, so long as we can identify eigenvalues with points on the BPS E-brane. 
We can conjecture that the string bits should interact in some way for this to happen, and that might be traced to the commutator squared terms in the SYM action. Indeed, these commutator terms 
also give rise to the interactions in the matrix string model \cite{DVV} \footnote{ Remember that the strings in the DVV model are  associated to field configurations on a circle with commuting matrices locally, and that to restore string interactions one need to restore the non-abelian nature of the model. This is necessarily related to non-zero commutators costing finite energy. 
In our case massive strings are non-abelian perturbations of the BPS membrane. However the string bits of two different strings usually commute as they don't have any eigenvalues in common.}.

\section{Discussion}\label{sec:disc}

In this paper we have worked hard to provide a qualitative picture of how spacetime geometry and locality appear from first principles in the the strong 't Hooft coupling limit of large $N$ limit of ${\cal N}=4 $ SYM theory. We have done this in a simplified setting where we could consider BPS geometries in supergravity, and tie them to BPS configurations in the field theory dual.

We studied the problem in half BPS geometries first, and then worked our way through the problem of classifying quarter BPS states and $1/8$ BPS states. It turns out that in all these cases the effective classical dynamics reduces to a system of $N$ 
non-relativistic particles associated to eigenvalues of various commuting matrices moving in
an even dimensional phase space. Here the phase space symplectic form serves as a strong magnetic field, so that one is studying a system with similar characteristics to a quantum hall
problem of particles in the lowest Landau level with some confining potential pushing all particles to the origin . 

One of the main results in this paper is that these particles are generally not free. They have
repulsive interactions (associated sometimes only to Fermi statistics) which arise from non-trivial measure terms in the reduction to eigenvalues. Because of the effective repulsive interactions, the ground state of the system forms a non-trivial geometry in phase space, which is large in units of $\hbar$. This is called the E-brane.
The shape of this E-brane can be obtained by a saddle point approximation of the many-body particle wave function. This requires some coarse-graining so that we replace this saddle point
problem by an effective statistical mechanics problem, with individual particles replaced by 
densities in phase space and excitations of the system by collective dynamics of the 
densities.

In some sense, in the ground state these particles are strongly interacting and the 
collective dynamics provides a much better description of the dynamics than the microscopic description of the system.

This gives us some geometric data which one might call a hydrodynamical description of the
dynamics of the system. We were able to show how coherent states 
associated to turning on gravitational waves on the AdS geometry work similarly in the hydrodynamical description of the BPS states, and deform the shape of the saddle point E-brane. This was done by using the AdS/CFT dictionary between 
gravitons and traces \cite{AdSCFT2}. Indeed, these shape deformations coincide with certain slices of the deformed $AdS_5\times S^5$ geometry, and suggests that in these cases we 
can identify the geometry we obtain by our coarse graining method with a slice of the full $AdS$ geometry. One should then be able to reconstruct the full $AdS$ deformed geometry by some procedure. In the work of Lin, Lunin and Maldacena \cite{LLM} for half BPS geometries, this involves solving
a differential equation with boundary conditions given by the hydrodynamic description of the states. In our case we conjecture that it works similarly for quarter BPS states, while for $1/8$ BPS states it probably requires a different method to deduce the ten dimensional geometry.

We were also able to show how the hydrodynamic geometry of the E-brane
can suffer topology changing transitions by condensing various defects which we associate to D-branes in the AdS geometry. These involve collective rearrangements of the 
non-relativistic particles.
One can speculate at this point that the topology of spacetime is a product of these collective rearrangements and groupings of particles alone. Since we have only finitely many particles, this means that we should only get finitely many possible groupings between them, and the internal organization of these groupings can only produce finitely many different shapes before the particles are so separated from each other that their collective dynamics loses 
it's meaning. If this is true, then this would have very profound consequences for quantum gravity.
Since this description  is in some sense a combinatorial problem with $N$ objects, 
the number of topologies will be bound by some function of $N$. If we think of 
topologies given by a very coarse description which only depends on the groupings of eigenvalues (and the number of eigenvalues on each grouping as a topological number), then the number of topologies associated to the large $N$ theory is 
roughly equal to the number of partitions of $N$, which grows exponentially with $N$.
If we look more closely, the geometric organization of the eigenvalues matters, and depending
on the amount of supersymmetry these can become more complicated, so the number is certainly larger. In any case, at the level of quantum gravity, if we need to sum over different topologies, there is a non-perturbative cutoff on the number of topologies that we can consider.
Indeed, we get that the number of BPS topologies $T$ is of order
\begin{equation}
\exp( a N) < T < f(N)
\end{equation}
Presumably $f(N) < \exp (b N^2)$ if we count just matrix degrees of freedom.
 This might not be too different from studying random graphs.

The important point is that the gravitational constant in five dimensions 
is of order $G\sim 1/N^2$, so that the number of topologies is bounded by $\exp(b/G)$.
We will call this bound the {\em topological exclusion principle}, similar to how one discusses other bounds on AdS/CFT for two dimensional CFT  as a stringy exclusion principle 
\cite{excl}.

This is important from many points of view. First, it suggests that topological features have a
minimum size in gravity. Presumably topology makes no sense at distances shorter than the Planck scale. Instead these features dissolve into the background and might become gravitational waves, or other low energy excitations, as discussed in \cite{LLM}.
This property also seems to be relevant in the study of the landscape \cite{MDtalk}.

 Even more important, the more complicated a topology one wants to construct,
  one needs to use a lot more pieces to build it,  and this implies that we need a lot 
  more information of how the system is configured. To store this information one would want to apply the holographic principle, and this requires 
  that complicated local topologies are bounded by large areas in the gravity theory.
  
This also seems to be due to the repulsive interactions between our non-relativistic 
particles. In general the more particles we try to put together, let's say $m$, 
 the larger their geometry gets, since it grows at least like $\sqrt m$ in our examples.

Going beyond BPS geometry, we have argued for the origin of massive string states as involving turning on non-trivial commutators. We needed to work at least in the quarter BPS geometries to understand how this might work in more detail.
We have  argued that this produces a picture where string bits join pairs of eigenvalues, and they end up being local on the geometry that these objects produce, namely a slice of $AdS_5\times S^5$
or some other semiclassical spacetime geometry. This is because for this amount of supersymmetry the  eigenvalues end up forming a very thin shell. The string bits end up being small because of energy and entropy considerations: long string bits connecting two well separated eigenvalues have a similar energy to a lot of little string bits connecting many
intermediate eigenvalues along a path on the geometry. These bits interact producing mixing between these states. Since this second class with a lot of small string bits has a lot more entropy,  mixing between all these states would be dominated by those that have large 
entropy, which shows that string states are described by curves on (a slice of) the spacetime geometry. 

Many of the arguments we have given in this paper are qualitative in nature, but they suggest a very natural setup in which to describe the emergence of spacetime geometry. A lot of the tests that we performed showed that we had good reason to believe that we were reproducing the geometry of gravitons and D-branes (giant gravitons) matching exactly the spacetime description. It is tempting to speculate that after the dynamics of the BPS configurations is taken care of properly, we will indeed reproduce the full geometry of an asymptotic $AdS_5\times S^5$ BPS supergravity configuration.

The fact that we seem to understand locality in AdS by understanding how geometric concepts 
becomes important to describe the CFT dynamics is an important step towards clarifying the details of the AdS/CFT dictionary. In particular, it suggests that we can prepare the system 
in the CFT so that the dual geometric object will collapse and form a black hole. This is currently under investigation \cite{BRV}.

We have also shown that the notion of geometry is reconstructed from the wave function 
of the Universe, which in this case can be understood as an initial condition on the dynamics. 
This realization of the notion of geometry is what we would call on-shell geometry. It is interesting to explore if off-shell dynamics in the CFT dual can be related to 
off-shell dynamics in the gravity theory or not.

\section*{ Acknowledgments }

I am very grateful to V. Balasubramanian, R. Corrado, D. Correa, S. Giddings,  D. Gross, J. Hartle, C. Herzog, P. Horava,  G. Horowitz, O. Lunin, J. Maldacena, L. Maoz, S. Pinansky, J. Polchinski, H. Reall, R. Roiban,C. Romelsberger,  M. Spradlin,  N. Seiberg, S. Vazquez, E. Verlinde, A. Volovich for many discussions related to this work. Work supported in part by a DOE outstanding Junior Investigator award, under grant DE-FG02-91ER40618.

\appendix

\section{Perturbative ${\cal N}=4 $ SYM on $S^3$}\label{sec:appa}

\subsection{The operator state correspondence}

The ${\cal N}=4$ SYM theory is a conformal field theory. This mean that there is an operator state corespondence of the following form. A gauge invariant local operator inserted at the origin  in Euclidean $\BR^4$, ${\cal O}(0)$ can be associated to a state for the SYM theory compactified
on an $S^3$. This arises from the fact that the metric on $\BR^4/0$ is conformally equivalent to the
metric on Euclidean $S^3\times \BR$. This metric is given explicitly in spherical coordinates as follows
\begin{equation}
ds^2 = r^2( dr^2/r^2 +d\Omega_3^2)
\end{equation}

 One then analytically continues this Euclidean manifold so that
time runs along $\BR$. One identifies the time coordinate with $dt \sim i dr/r$, so that $t\sim \log(r)$

The Hamiltonian associated to time evolution evolution is $\partial_t\sim r\partial_r = \Delta$, which is the generator of dilatations around the origin. This is how one identifies the generator of dilatations with
time evolution.

At the free field theory level, we can classify all local gauge invariant operators as polynomials in the fields and their derivative. One has to mod out by the equations of motion to do this. Terms involving the equations of motion 
will generically be zero on correlators, except perhaps for contact terms.

In any case, we can make a list of the symbols we can use. We can use fields and their derivatives.
The spectrum of ${\cal N}=4 $ SYM theory consists of six real scalars $X^i$, four Weyl fermions
$\psi^j_{\alpha}$, and the spin one gauge field $A_\mu$ in the adjoint of $SU(N)$.

Their quantum numbers are given in the following table. There is an $SO(6)$ R-symmetry group. The cartan basis is given by $J_1, J_2, J_3$. We also use $J=J_1+J_2+J_3$. We use complex combinations of the 
$X^i$, $\phi^i$, $\bar\phi^i$, for $i=1, 2, 3$. We also classify the spin under the $SO(4)$ rotations of the Euclidean plane.

\begin{tabular}{|c|c|c|c|c|c|}
 \hline
Field & Spin &$ \Delta$ & $J_1$ & $J$\\
\hline $\phi^1$ &$(0,0)$& 1 & 1 & 1\\
$\phi^{2,3}$&$(0,0)$&1&0&1\\
$\bar \phi^1$ & $(0,0)$&1&-1 &-1\\
$\bar\phi^{2,3}$&$(0,0)$&1&0&-1\\
$\psi^{\alpha}$&$(1/2,0)$&3/2&$\pm 1/2$&$ 3\times - 1/2$, $3/2$\\
$\bar \psi^{\alpha}$&$(0,1/2)$&3/2&$\pm 1/2$&$3\times 1/2 $, $ - 3/2$\\
$F^{\mu\nu}$ & $(1,0)\oplus (0,1)$ &2 &0&0\\
\hline
\end{tabular}

The derivatives have $\Delta=1$,  spin $(1/2,1/2)$ and $R$ charge zero. The supersymemtry generators have $J$ charge given by $\pm 1/2, \pm 3/2$ and conformal weight $\pm 1/2$.

We can also quantize the free field theory on the $S^3$ and decompose all fields in terms of 
spherical harmonics decomposition. 
For the scalars, they decompose as $(n/2,n/2)$ representation of $SO(4)$, for fermions, they decompose
as $(n/2+1/2,n/2)$ and as $(n/2,n/2+1/2)$, and for the vectors they do as $(n/2+1,n/2)\oplus(n/2,n/2+1)$.
For each spherical harmonic there is both a raising operator $a^\dagger$ and a lowering operator $a$.

It is easy to match the spherical harmonic decomposition for the scalar with operators as follows
$\phi_{n,n} \sim \partial^n\phi(0)$. A similar reasoning follows for spinors and tensors. We are to identify the insertion of the corresponding  operator at the origin with the corresponding creation operator of the quantum field with the given symmetries. 
In the lagrangian of ${\cal N}=4$ SYM on $S^3$ there is a conformal coupling of the scalars to the metric on $S^3$. It is easy to show that this makes the energies of the corresponding quanta integer valued, and one can match the table given above exactly.

The operators that satisfy $\Delta= J_1$ are given by just one complex scalar: the s-wave of the field $\phi^1$ on $S^3$. The operators that satisfy $\Delta=J$ are the three fields $\phi^{1,2,3}$ and two  spin polarizations of the fields $\psi$. Notice that these two polarizations have spin $0$ for one of the subgroups of the $SU(2)$ rotations about the origin.

\subsection{One loop calculations in $1/8$ BPS configurations }

The scalar lagrangian of ${\cal N}=4 $ SYM on $S^3$ is given by

\begin{equation}
L = \frac 1{g_{YM}^2}\int d^3 \Omega dt \frac 12 (D_\mu X^i)^2- \frac 14 ([X^i,X_j])^2- \frac 12 (X^i)^2
\end{equation}
We want to have $1/8$ BPS configurations that satisfy $\Delta=J$ classically. This requires that $X$ be independent of the angles on $S^3$. Since $\Delta $and $J$ commute, we can diagonalize them as operators simultaneously, and twist the Hamiltonian to be equal to $H=\Delta-J$. With this definition of $H$, the supersymmetry algebra takes the form $\sum \{Q_i,Q_i\} \sim H$ for four real supercharges which turn out not to commute with $H$. However,  $H\ket\psi=0$ still implies that
the states satisfy $Q\ket\psi=0$ for all four $Q$. These $Q$ have $J$ charge equal to 
$\pm 3/2$, and conformal dimension equal to $\pm 1/2$. However, they commute with  
a half BPS hamiltonian $\tilde H= \Delta -J_1$. 

Classically this requires all $X$ to be commuting matrices as argued in section \ref{sec:matrix}. Moreover, one can see that the $X$ have a specific time dependence with respect to $\Delta$, so that
seen as a classical system we end up with three pairs of canonically conjugate variables. This is because the value of the matrices $X$ determines $\dot X$. This can be understood as a magnetic effect on the coordinates $X$. If we use the Hamiltonian $H$ instead, the variables $X$ are time independent.
Now, because one can also have fermionic ground states, one can in principle only guarantee that
the Witten index of the system is preserved. This might lift some configurations by quantum effects.
We want to see that that does not happen.

To do a calculation of leading order quantum effects, we split the dynamics into zero modes associated to the eigenvalues of the matrices $X$ which are treated as collective coordinates, and we integrate everything else out, by using the quadratic lagrangian on the modes that are not zero modes, with the zero modes treated exactly (and given by constant values).

Treating $X$ exactly and giving it constant values means that we can diagonalize all the $X$
simultaneosly. For each eigenvalue we have six real coordinates $\vec \lambda_i$, for $i=1,\dots N$. We will 
assume now that we are in a generic situation where all the $\vec \lambda_i$ are distinct.
This breaks the gauge symmetry from $U(N)$ to $U(1)^{N}$.

We now decompose all fields on the $S^3$ as spherical harmonics, and we also split them according to their charges under $U(1)^N$. This is the same as splitting it according to components that join two eigenvalues $M^i_j$.

For the scalar fields, let $\delta X^i_j$  be a mode we are integrating out. By the previous section, it decomposes as the $(n/2,n/2)$ representation of the $SO(4)$ group for all values of $n$.

The quadratic piece of the Hamiltonian has three pieces. The usual kinetic term and mass term of the free field theory (which includes the gradient terms on the $S^3$). If we have twisted the Hamiltonian by $J$, the $J$ charge of the field modifies the Hamiltonian for the quadratic modes by counting the J-charge of the configuration.
An additional contribution to the mass arises from the commutator squared terms, which explicitly goes like
\begin{equation}
\tr([X,\delta X][X,\delta X]) = ( (X_{ii}-X_{jj})^2 \delta X^i_j\cdot \delta X^j_i)-(X_{ii}-X_{jj})\cdot \delta X^i_j
(X_{ii}-X_{jj})\cdot \delta X^{j}_i
\end{equation}
In the notation above the dots indicate the vector dot products.

Now, we can choose the gauge fixing of the gauge theory on $S^3$ so that the second term with 
the dot product of $X$ and $\delta X$ vanishes. For generic $i,j$, this removes one scalar 
polarization, which has $J$ charge equal to one: it is proportional to a $SU(N)$ rotation of the $X$.
However, it turns out to be more convenient to use a gauge fixing procedure like the Feynman gauge, where we eliminate this term in the action by introducing ghosts of the same mass than this polarization (this type of gauge fixing was used in \cite{BC}). 

 All of them are in the $(n/2,n/2)$ representation of $SU(2)\times SU(2)$. Each of these will have a total of $(n+1)^2$ degrees of freedom, and there are six degrees of freedom in total. 

For the gauge bosons, we will have fields in the $(n/2+1,n/2)$ and $(n/2-1,n/2)$ representation of $SO(4)$. We should look to match states which have the same $SU(2)$ rotational charge that the scalars. This is because the unbroken supersymmetries do not transform under one of the $SU(2)$ rotation groups inside $SO(4)$.
This gives us an additional degeneracy of $(n+1)(n+3)$ or $(n-1)(n+1)$ polarizations from spin degeneracy.
 
Finally, for the fermions we will have $4$ physical polarizations in the $(n/2+1/2,n/2)$ and
another four in the $(n/2-1/2,n/2)$ for a total degeneracy of states of order $4(n+2)(n+1)+4(n)(n+1)$.

One can easily check that the total number of boson and fermion polarizations for each value of $n$ are equal to each other. Indeed, they should be related by the unbroken supersymmetries of the configuration, as all one particle excitations should be classified by the unbroken symmetry of the configuration. Indeed, all of them that are grouped in the same multiplet have to have the same value of $H$.
Because the $J$ charge of one of these multiplets adds to zero, it's easy to see that the contribution to either $\Delta$ or $H$ of these sates adds to zero.

This shows that to one loop order the energy of the moduli space of vacua we have been considering does not get lifted by quantum corrections. This proof covers the $1/8$ BPS bosonic configurations.
In principle, this does not imply that the effective action is not modified. This result is to be considered
as an on-shell calculation because we are requiring the equations of motion to be satisfied (this is what $X$ constant does for us).


\begin{thebibliography}{99}

\bibitem{QCD}
  H.~D.~Politzer,
  ``Reliable Perturbative Results For Strong Interactions?,''
  Phys.\ Rev.\ Lett.\  {\bf 30}, 1346 (1973).
  %%CITATION = PRLTA,30,1346;%%
  D.~J.~Gross and F.~Wilczek,
  ``Ultraviolet Behavior Of Non-Abelian Gauge Theories,''
  Phys.\ Rev.\ Lett.\  {\bf 30}, 1343 (1973).
  %%CITATION = PRLTA,30,1343;%%
  D.~J.~Gross and F.~Wilczek,
  ``Asymptotically Free Gauge Theories. 1,''
  Phys.\ Rev.\ D {\bf 8}, 3633 (1973).
  %%CITATION = PHRVA,D8,3633;%%%
    D.~J.~Gross and F.~Wilczek,
  ``Asymptotically Free Gauge Theories. 2,''
  Phys.\ Rev.\ D {\bf 9}, 980 (1974).
  %%CITATION = PHRVA,D9,980;%%


%\cite{'tHooft:1973jz}
\bibitem{largeN}
G.~'t Hooft,
``A Planar Diagram Theory For Strong Interactions,''
Nucl.\ Phys.\ B {\bf 72}, 461 (1974).
%%CITATION = NUPHA,B72,461;%%

%\cite{Witten:1979kh}
\bibitem{Witbaryons}
  E.~Witten,
  ``Baryons In The 1/N Expansion,''
  Nucl.\ Phys.\ B {\bf 160}, 57 (1979).
  %%CITATION = NUPHA,B160,57;%%
  
  
  %\cite{Maldacena:1997re}
\bibitem{M}
  J.~M.~Maldacena,
 ``The large N limit of superconformal field theories and supergravity,''
  Adv.\ Theor.\ Math.\ Phys.\  {\bf 2}, 231 (1998)
  [Int.\ J.\ Theor.\ Phys.\  {\bf 38}, 1113 (1999)]
  [arXiv:hep-th/9711200].
  %%CITATION = HEP-TH 9711200;%%
  


\bibitem{Holographic}
  G.~'t Hooft,
``Dimensional reduction in quantum gravity,''
  arXiv:gr-qc/9310026.
  %%CITATION = GR-QC 9310026;%%
  L.~Susskind,
``The World as a hologram,''
  J.\ Math.\ Phys.\  {\bf 36}, 6377 (1995)
  [arXiv:hep-th/9409089].
  %%CITATION = HEP-TH 9409089;%%  
  

\bibitem{Bousso}
  R.~Bousso,
  ``A Covariant Entropy Conjecture,''
  JHEP {\bf 9907}, 004 (1999)
  [arXiv:hep-th/9905177].
  %%CITATION = HEP-TH 9905177;%%
  R.~Bousso,
  ``The holographic principle,''
  Rev.\ Mod.\ Phys.\  {\bf 74}, 825 (2002)
  [arXiv:hep-th/0203101].
  %%CITATION = HEP-TH 0203101;%%

 %\cite{Polyakov:1997tj}
\bibitem{Polyakov}
  A.~M.~Polyakov,
``String theory and quark confinement,''
  Nucl.\ Phys.\ Proc.\ Suppl.\  {\bf 68}, 1 (1998)
  [arXiv:hep-th/9711002].
  %%CITATION = HEP-TH 9711002;%%
  A.~M.~Polyakov,
  ``The wall of the cave,''
  Int.\ J.\ Mod.\ Phys.\ A {\bf 14}, 645 (1999)
  [arXiv:hep-th/9809057].
  %%CITATION = HEP-TH 9809057;%%

  
  \bibitem{Mikh}
  A.~Mikhailov,
  ``Giant gravitons from holomorphic surfaces,''
  JHEP {\bf 0011}, 027 (2000)
  [arXiv:hep-th/0010206].
  %%CITATION = HEP-TH 0010206;%%
  


%\cite{Corley:2001zk}
\bibitem{CJR}
S.~Corley, A.~Jevicki and S.~Ramgoolam,
``Exact correlators of giant gravitons from dual N = 4 SYM theory,''
Adv.\ Theor.\ Math.\ Phys.\  {\bf 5}, 809 (2002)
[arXiv:hep-th/0111222].
%%CITATION = HEP-TH 0111222;%%
  
 
%\cite{Kristjansen:2002bb}
\bibitem{J2N}
  C.~Kristjansen, J.~Plefka, G.~W.~Semenoff and M.~Staudacher,
  ``A new double-scaling limit of N = 4 super Yang-Mills theory and PP-wave
  strings,''
  Nucl.\ Phys.\ B {\bf 643}, 3 (2002)
  [arXiv:hep-th/0205033].
  %%CITATION = HEP-TH 0205033;%%
  N.~R.~Constable, D.~Z.~Freedman, M.~Headrick, S.~Minwalla, L.~Motl, A.~Postnikov and W.~Skiba,
``PP-wave string interactions from perturbative Yang-Mills theory,''
  JHEP {\bf 0207}, 017 (2002)
  [arXiv:hep-th/0205089].
  %%CITATION = HEP-TH 0205089;%%

%\cite{Berenstein:2002sa}
\bibitem{BN}
  D.~Berenstein and H.~Nastase,
 ``On lightcone string field theory from super Yang-Mills and holography,''
  arXiv:hep-th/0205048.
  %%CITATION = HEP-TH 0205048;%%
 
 %\cite{Berenstein:2002jq}
\bibitem{BMN}
  D.~Berenstein, J.~M.~Maldacena and H.~Nastase,
 ``Strings in flat space and pp waves from N = 4 super Yang Mills,''
  JHEP {\bf 0204}, 013 (2002)
  [arXiv:hep-th/0202021].
  %%CITATION = HEP-TH 0202021;%%
  
  
%\cite{Berenstein:2004kk}
\bibitem{toyads}
D.~Berenstein,
``A toy model for the AdS/CFT correspondence,''
JHEP {\bf 0407}, 018 (2004)
[arXiv:hep-th/0403110].
%%CITATION = HEP-TH 0403110;%%


%\cite{Brezin:1977sv}
\bibitem{BIPZ}
E.~Brezin, C.~Itzykson, G.~Parisi and J.~B.~Zuber,
``Planar Diagrams,''
Commun.\ Math.\ Phys.\  {\bf 59}, 35 (1978).
%%CITATION = CMPHA,59,35;%%



%\cite{Lin:2004nb}
\bibitem{LLM}
  H.~Lin, O.~Lunin and J.~Maldacena,
  ``Bubbling AdS space and 1/2 BPS geometries,''
  JHEP {\bf 0410}, 025 (2004)
  [arXiv:hep-th/0409174].
  %%CITATION = HEP-TH 0409174;%%


%\cite{Lee:1998bx}
\bibitem{LMRS}
  S.~M.~Lee, S.~Minwalla, M.~Rangamani and N.~Seiberg,
  ``Three-point functions of chiral operators in D = 4, N = 4 SYM at  large
  N,''
  Adv.\ Theor.\ Math.\ Phys.\  {\bf 2}, 697 (1998)
  [arXiv:hep-th/9806074].
  %%CITATION = HEP-TH 9806074;%%

%\cite{Laughlin:1983fy}
\bibitem{Laughlin}
R.~B.~Laughlin,
``Anomalous Quantum Hall Effect: An Incompressible Quantum Fluid With
Fractionally Charged Excitations,''
Phys.\ Rev.\ Lett.\  {\bf 50} (1983) 1395.
%%CITATION = PRLTA,50,1395;%%

\bibitem{QHE}
``The Quantum Hall effect'', edited by R.E. Prange and S. M. Girvin, 
Springer Verlag, NY, 1987.


\bibitem{AdSCFT2}
 S.~S.~Gubser, I.~R.~Klebanov and A.~M.~Polyakov,
``Gauge theory correlators from non-critical string theory,''
  Phys.\ Lett.\ B {\bf 428}, 105 (1998)
  [arXiv:hep-th/9802109].
  %%CITATION = HEP-TH 9802109;%%
  E.~Witten,
 ``Anti-de Sitter space and holography,''
  Adv.\ Theor.\ Math.\ Phys.\  {\bf 2}, 253 (1998)
  [arXiv:hep-th/9802150].
  %%CITATION = HEP-TH 9802150;%%

%\cite{Stone:1990iw}
\bibitem{Stoneedge}
M.~Stone,
``Edge Waves In The Quantum Hall Effect,''
Annals Phys.\  {\bf 207}, 38 (1991).
%%CITATION = APNYA,207,38;%%



\bibitem{BBNS}
  V.~Balasubramanian, M.~Berkooz, A.~Naqvi and M.~J.~Strassler,
``Giant gravitons in conformal field theory,''
  JHEP {\bf 0204}, 034 (2002)
  [arXiv:hep-th/0107119].
  %%CITATION = HEP-TH 0107119;%%



\bibitem{Metal}
  R.~C.~Myers and O.~Tafjord,
``Superstars and giant gravitons,''
  JHEP {\bf 0111}, 009 (2001)
  [arXiv:hep-th/0109127].
  %%CITATION = HEP-TH 0109127;%%
  
\bibitem{MiL}
  G.~Milanesi and M.~O'Loughlin,
 ``Singularities and closed time-like curves in type IIB 1/2 BPS geometries,''
  arXiv:hep-th/0507056.
  %%CITATION = HEP-TH 0507056;%%
%\cite{Hartle:1983ai}
\bibitem{HH}
  J.~B.~Hartle and S.~W.~Hawking,
``Wave Function Of The Universe,''
  Phys.\ Rev.\ D {\bf 28}, 2960 (1983).
  %%CITATION = PHRVA,D28,2960;%%

%\cite{Ooguri:2005vr}
\bibitem{BPStop}
  H.~Ooguri, C.~Vafa and E.~Verlinde,
  ``Hartle-Hawking wave-function for flux compactifications,''
  arXiv:hep-th/0502211.
  %%CITATION = HEP-TH 0502211;%%
  M.~Aganagic, A.~Neitzke and C.~Vafa,
  ``BPS microstates and the open topological string wave function,''
  arXiv:hep-th/0504054.
  %%CITATION = HEP-TH 0504054;%%
  R.~Dijkgraaf, R.~Gopakumar, H.~Ooguri and C.~Vafa,
``Baby universes in string theory,''
  arXiv:hep-th/0504221.
  %%CITATION = HEP-TH 0504221;%%
    S.~Gukov, K.~Saraikin and C.~Vafa,
 ``A stringy wave function for an S**3 cosmology,''
  arXiv:hep-th/0505204.
  %%CITATION = HEP-TH 0505204;%%
  
  
%\cite{Klebanov:1991qa}
\bibitem{kleb}
I.~R.~Klebanov,
``String theory in two-dimensions,''
arXiv:hep-th/9108019.
%%CITATION = HEP-TH 9108019;%%

\bibitem{MTV}
  J.~McGreevy, J.~Teschner and H.~Verlinde,
  ``Classical and quantum D-branes in 2D string theory,''
  JHEP {\bf 0401}, 039 (2004)
  [arXiv:hep-th/0305194].
  %%CITATION = HEP-TH 0305194;%%


\bibitem{MMSS}
  J.~Maldacena, G.~W.~Moore, N.~Seiberg and D.~Shih,
``Exact vs. semiclassical target space of the minimal string,''
  JHEP {\bf 0410}, 020 (2004)
  [arXiv:hep-th/0408039].
  %%CITATION = HEP-TH 0408039;%%


 %\cite{Mandal:2005wv}
\bibitem{Collect}
  G.~Mandal,
  ``Fermions from half-BPS supergravity,''
  arXiv:hep-th/0502104.
  %%CITATION = HEP-TH 0502104;%%
  L.~Grant, L.~Maoz, J.~Marsano, K.~Papadodimas and V.~S.~Rychkov,
  ``Minisuperspace quantization of 'bubbling AdS' and free fermion droplets,''
  arXiv:hep-th/0505079.
  %%CITATION = HEP-TH 0505079;%%
  A.~Dhar,
 ``Bosonization of non-relativstic fermions in 2-dimensions and collective
  field theory,''
  arXiv:hep-th/0505084.
  %%CITATION = HEP-TH 0505084;%%



%\cite{McGreevy:2000cw}
\bibitem{gg}
  J.~McGreevy, L.~Susskind and N.~Toumbas,
``Invasion of the giant gravitons from anti-de Sitter space,''
  JHEP {\bf 0006}, 008 (2000)
  [arXiv:hep-th/0003075].
  %%CITATION = HEP-TH 0003075;%%
A.~Hashimoto, S.~Hirano and N.~Itzhaki,
  ``Large branes in AdS and their field theory dual,''
  JHEP {\bf 0008}, 051 (2000)
  [arXiv:hep-th/0008016].
  %%CITATION = HEP-TH 0008016;%%
 M.~T.~Grisaru, R.~C.~Myers and O.~Tafjord,
 ``SUSY and Goliath,''
  JHEP {\bf 0008}, 040 (2000)
  [arXiv:hep-th/0008015].
  %%CITATION = HEP-TH 0008015;%%
  
  \bibitem{giants}
   V.~Balasubramanian, M.~x.~Huang, T.~S.~Levi and A.~Naqvi,
  ``Open strings from N = 4 super Yang-Mills,''
  JHEP {\bf 0208}, 037 (2002)
  [arXiv:hep-th/0204196].
  %%CITATION = HEP-TH 0204196;%%
O.~Aharony, Y.~E.~Antebi, M.~Berkooz and R.~Fishman,
``'Holey sheets': Pfaffians and subdeterminants as D-brane operators in large
N gauge theories,''
JHEP {\bf 0212}, 069 (2002)
[arXiv:hep-th/0211152].
%%CITATION = HEP-TH 0211152;%%
D.~Berenstein,
``Shape and holography: Studies of dual operators to giant gravitons,''
Nucl.\ Phys.\ B {\bf 675}, 179 (2003)
[arXiv:hep-th/0306090].
%%CITATION = HEP-TH 0306090;%%
 V.~Balasubramanian, D.~Berenstein, B.~Feng and M.~x.~Huang,
  ``D-branes in Yang-Mills theory and emergent gauge symmetry,''
  JHEP {\bf 0503}, 006 (2005)
  [arXiv:hep-th/0411205].
  %%CITATION = HEP-TH 0411205;%%
  D.~Berenstein and S.~E.~Vazquez,
  ``Integrable open spin chains from giant gravitons,''
  arXiv:hep-th/0501078.
  %%CITATION = HEP-TH 0501078;%%
  D.~Berenstein, D.~H.~Correa and S.~E.~Vazquez,
``Quantizing open spin chains with variable length: An example from giant
gravitons,''
  arXiv:hep-th/0502172.
  %%CITATION = HEP-TH 0502172;%%



\bibitem{CKS}
  M.~M.~Caldarelli, D.~Klemm and P.~J.~Silva,
  ``Chronology protection in anti-de Sitter,''
  arXiv:hep-th/0411203.
  %%CITATION = HEP-TH 0411203;%%



\bibitem{HS}
  P.~Horava and P.~G.~Shepard,
``Topology changing transitions in bubbling geometries,''
  JHEP {\bf 0502}, 063 (2005)
  [arXiv:hep-th/0502127].
  %%CITATION = HEP-TH 0502127;%%
  

\bibitem{BJS}
  V.~Balasubramanian, V.~Jejjala and J.~Simon,
  ``The library of Babel,''
  arXiv:hep-th/0505123.
  %%CITATION = HEP-TH 0505123;%%
   V.~Balasubramanian, Jan de Boer, V.~Jejjala and J.~Simon,
   ``The library of Babel: On he origin of gravitational thermodynamics,"
   {\em to appear}

\bibitem{TY}
  Y.~Takayama and K.~Yoshida,
  ``Bubbling 1/2 BPS geometries and Penrose limits,''
  arXiv:hep-th/0503057.
  %%CITATION = HEP-TH 0503057;%%
  
  
  
 
%\cite{Mukhi:2005cv}
\bibitem{Mukhi}
  S.~Mukhi and M.~Smedback,
 ``Bubbling orientifolds,''
  arXiv:hep-th/0506059.
  %%CITATION = HEP-TH 0506059;%% 
  
  \bibitem{TT}
  Y.~Takayama and A.~Tsuchiya,
``Complex Matrix Model and Fermion Phase Space for Bubbling AdS Geometries,''
  arXiv:hep-th/0507070.
  %%CITATION = HEP-TH 0507070;%%
  
  
  %\cite{Itzhaki:2004te}
\bibitem{IMc}
N.~Itzhaki and J.~McGreevy,
``The large N harmonic oscillator as a string theory,''
arXiv:hep-th/0408180.
%%CITATION = HEP-TH 0408180;%%
  

  


%\cite{Ghodsi:2005ks}
\bibitem{Ghosi}
  A.~Ghodsi, A.~E.~Mosaffa, O.~Saremi and M.~M.~Sheikh-Jabbari,
  ``LLL vs. LLM: Half BPS sector of N = 4 SYM equals to quantum Hall system,''
  arXiv:hep-th/0505129.
  %%CITATION = HEP-TH 0505129;%%

 \bibitem{DJR}
  A.~Donos, A.~Jevicki and J.~P.~Rodrigues,
 ``Matrix model maps in AdS/CFT,''
  arXiv:hep-th/0507124.
  %%CITATION = HEP-TH 0507124;%%


%\cite{Cachazo:2002ry}
\bibitem{CDSW}
  F.~Cachazo, M.~R.~Douglas, N.~Seiberg and E.~Witten,
  ``Chiral rings and anomalies in supersymmetric gauge theory,''
  JHEP {\bf 0212}, 071 (2002)
  [arXiv:hep-th/0211170].
  %%CITATION = HEP-TH 0211170;%%

  %\cite{Banks:1996vh}
\bibitem{BFSS}
  T.~Banks, W.~Fischler, S.~H.~Shenker and L.~Susskind,
 ``M theory as a matrix model: A conjecture,''
  Phys.\ Rev.\ D {\bf 55}, 5112 (1997)
  [arXiv:hep-th/9610043].
  %%CITATION = HEP-TH 9610043;%%
  
\bibitem{quarterBPS}
  A.~V.~Ryzhov,
 ``Quarter BPS operators in N = 4 SYM,''
  JHEP {\bf 0111}, 046 (2001)
  [arXiv:hep-th/0109064].
  %%CITATION = HEP-TH 0109064;%%
  E.~D'Hoker and A.~V.~Ryzhov,
 ``Three-point functions of quarter BPS operators in N = 4 SYM,''
  JHEP {\bf 0202}, 047 (2002)
  [arXiv:hep-th/0109065].
  %%CITATION = HEP-TH 0109065;%%
  E.~D'Hoker, P.~Heslop, P.~Howe and A.~V.~Ryzhov,
 ``Systematics of quarter BPS operators in N = 4 SYM,''
  JHEP {\bf 0304}, 038 (2003)
  [arXiv:hep-th/0301104].
  %%CITATION = HEP-TH 0301104;%%

%\cite{Beisert:2003yb}
\bibitem{Beietal}
  J.~A.~Minahan and K.~Zarembo,
  ``The Bethe-ansatz for N = 4 super Yang-Mills,''
  JHEP {\bf 0303}, 013 (2003)
  [arXiv:hep-th/0212208].
  %%CITATION = HEP-TH 0212208;%% 
   N.~Beisert,
  ``The complete one-loop dilatation operator of N = 4 super Yang-Mills
  theory,''
  Nucl.\ Phys.\ B {\bf 676}, 3 (2004)
  [arXiv:hep-th/0307015].
  %%CITATION = HEP-TH 0307015;%%
  N.~Beisert and M.~Staudacher,
``The N = 4 SYM integrable super spin chain,''
  Nucl.\ Phys.\ B {\bf 670}, 439 (2003)
  [arXiv:hep-th/0307042].
  %%CITATION = HEP-TH 0307042;%%
  N.~Beisert,
``The su(2$|$3) dynamic spin chain,''
  Nucl.\ Phys.\ B {\bf 682}, 487 (2004)
  [arXiv:hep-th/0310252].
  %%CITATION = HEP-TH 0310252;%%


\bibitem{Pol}
  J.~Polchinski,
 ``On the nonperturbative consistency of d = 2 string theory,''
  Phys.\ Rev.\ Lett.\  {\bf 74}, 638 (1995)
  [arXiv:hep-th/9409168].
  %%CITATION = HEP-TH 9409168;%%

\bibitem{Demel}
  R.~de Mello Koch and R.~Gwyn,
  ``Giant graviton correlators from dual SU(N) super Yang-Mills theory,''
  JHEP {\bf 0411}, 081 (2004)
  [arXiv:hep-th/0410236].
  %%CITATION = HEP-TH 0410236;%%


%\cite{Mathur:2003hj}
\bibitem{Mathur}
  S.~D.~Mathur, A.~Saxena and Y.~K.~Srivastava,
  ``Constructing 'hair' for the three charge hole,''
  Nucl.\ Phys.\ B {\bf 680}, 415 (2004)
  [arXiv:hep-th/0311092].
  %%CITATION = HEP-TH 0311092;%%
   S.~D.~Mathur,
``The fuzzball proposal for black holes: An elementary review,''
  arXiv:hep-th/0502050.
  %%CITATION = HEP-TH 0502050;%%

\bibitem{LMM}
  O.~Lunin, J.~Maldacena and L.~Maoz,
  ``Gravity solutions for the D1-D5 system with angular momentum,''
  arXiv:hep-th/0212210.
  %%CITATION = HEP-TH 0212210;%%

%\cite{Seiberg:1994rs}
\bibitem{SW}
  N.~Seiberg and E.~Witten,
``Electric - magnetic duality, monopole condensation, and confinement in N=2
supersymmetric Yang-Mills theory,''
  Nucl.\ Phys.\ B {\bf 426}, 19 (1994)
  [Erratum-ibid.\ B {\bf 430}, 485 (1994)]
  [arXiv:hep-th/9407087].
  %%CITATION = HEP-TH 9407087;%%


\bibitem{BC}
  D.~Berenstein and R.~Corrado,
 ``M(atrix)-theory in various dimensions,''
  Phys.\ Lett.\ B {\bf 406}, 37 (1997)
  [arXiv:hep-th/9702108].
  %%CITATION = HEP-TH 9702108;%%

%\cite{Taylor:1996ik}
\bibitem{WT}
  W.~I.~Taylor,
  ``D-brane field theory on compact spaces,''
  Phys.\ Lett.\ B {\bf 394}, 283 (1997)
  [arXiv:hep-th/9611042].
  %%CITATION = HEP-TH 9611042;%%
  
\bibitem{PE}
  H.~Elvang and J.~Polchinski,
  ``The quantum Hall effect on R**4,''
  arXiv:hep-th/0209104.
  %%CITATION = HEP-TH 0209104;%%
 D.~Karabali and V.~P.~Nair,
``The effective action for edge states in higher dimensional quantum Hall
 systems,''
  Nucl.\ Phys.\ B {\bf 679}, 427 (2004)
  [arXiv:hep-th/0307281].
  %%CITATION = HEP-TH 0307281;%%
 D.~Karabali and V.~P.~Nair,
  ``Edge states for quantum Hall droplets in higher dimensions and a
  generalized WZW model,''
  Nucl.\ Phys.\ B {\bf 697}, 513 (2004)
  [arXiv:hep-th/0403111].
  %%CITATION = HEP-TH 0403111;%%
  A.~P.~Polychronakos,
  ``Chiral actions from phase space (quantum Hall) droplets,''
  Nucl.\ Phys.\ B {\bf 705}, 457 (2005)
  [arXiv:hep-th/0408194].
  %%CITATION = HEP-TH 0408194;%%
  

\bibitem{Kor}
  S.~Kim and K.~Lee,
``BPS electromagnetic waves on giant gravitons,''
  arXiv:hep-th/0502007.
  %%CITATION = HEP-TH 0502007;%%
  
  
  \bibitem{DVV}
  R.~Dijkgraaf, E.~P.~Verlinde and H.~L.~Verlinde,
  ``Matrix string theory,''
  Nucl.\ Phys.\ B {\bf 500}, 43 (1997)
  [arXiv:hep-th/9703030].
  %%CITATION = HEP-TH 9703030;%%

  
  \bibitem{excl}
  J.~M.~Maldacena and A.~Strominger,
  ``AdS(3) black holes and a stringy exclusion principle,''
  JHEP {\bf 9812}, 005 (1998)
  [arXiv:hep-th/9804085].
  %%CITATION = HEP-TH 9804085;%%
  
  \bibitem{MDtalk}
  M.~R.~ Douglas,
  ``Is the number of string vacua finite?",  talk
  at strings 2005. 
  
  \bibitem{BRV}
  D.~Berenstein, H.~Reall, S. Vazquez,
  {\em Work in progress}
  







%%\cite{Polychronakos:2004es}
%\bibitem{Poly2}
%A.~P.~Polychronakos,
%``Chiral actions from phase space (quantum Hall) droplets,''
%arXiv:hep-th/0408194.
%%%CITATION = HEP-TH 0408194;%%










  
\end{thebibliography}
 \end{document}